\pdfoutput=1  

\documentclass[a4paper, final]{article} 

\usepackage[ruled]{algorithm2e}

\SetAlFnt{\small}
\SetAlCapFnt{\small}
\SetAlCapNameFnt{\small}
\SetAlCapHSkip{0pt}
\IncMargin{-\parindent}

\usepackage{hyphenat}
\usepackage[mode=text]{siunitx}
\usepackage{rotating}
\usepackage{subfigure}
\usepackage{tikz}
\usepackage{listings}
\usepackage{threeparttable}
\usepackage{gensymb}
\usepackage{multirow}
\usepackage{caption}
\usepackage[margin=3cm]{geometry} 

\newcommand{\io}{IO}
\newcommand{\ooo}{OOO}
\newcommand{\OOO}{OOO}
\newcommand{\degoal}{deGoal}
\newcommand{\monocol}[1]{\multicolumn{1}{#1}}

\newcommand{\lstCplainfileCustom}[1]
{
  \lstinputlisting[
    xleftmargin=15pt,
    language=C,
    numbers=left,
    stepnumber=1,
    basicstyle=\ttfamily\scriptsize,
    keywordstyle=\color{black}\ttfamily,
    commentstyle=\color{green!50!black},
    stringstyle=\color{red}\ttfamily,
    showstringspaces=false,
    breaklines=true,
    escapechar=\%,
    classoffset=1,
    emph={hotUF, coldUF, vectLen, pldStride},
    emphstyle=\itshape\color{red},
    classoffset=2,
    emph={int, for, if},
    emphstyle=\color{blue},
    classoffset=0
  ]{#1}
}

\begin{document}

\markboth{F. A. Endo et al.}{Pushing the Limits of Online Auto-tuning:
Machine Code Optimization in Short-Running Kernels}

\title{Pushing the Limits of Online Auto-tuning: Machine Code Optimization in
Short-Running Kernels\footnote{Extension of a Conference Paper published in the proceedings of MCSoC-16:
IEEE 10th International Symposium on Embedded Multicore/Many-core
Systems-on-Chip, Lyon, France, 2016.}}

\author{Fernando A. Endo \and Damien Courouss\'e \and Henri-Pierre Charles \\
Univ. Grenoble Alpes, F-38000 Grenoble, France \\ CEA, LIST, MINATEC Campus, F-38054 Grenoble, France
}

\date{}

\maketitle

\begin{abstract}
%

We propose an online auto-tuning approach for computing kernels.
Differently from existing online auto-tuners, which regenerate code with long
compilation chains from the source to the binary code, our approach consists on
deploying auto-tuning directly at the level of machine code generation. This
allows auto-tuning to pay off in very short-running applications. As a proof of
concept, our approach is demonstrated in two benchmarks, which execute during
hundreds of milliseconds to a few seconds only. In a CPU-bound kernel, the
 average speedups achieved are 1.10 to 1.58 depending on the target micro-architecture, up to 2.53 in the most favourable conditions
(all run-time overheads included).
In a memory-bound kernel, less favourable to our runtime auto-tuning optimizations, the  average speedups are 1.04 to 1.10, up to 1.30 in the best configuration.
Despite the short execution times of our benchmarks, the overhead of our runtime auto-tuning is between 0.2 and 4.2~\% only of the total application execution times.
By simulating the CPU-bound application in 11 different CPUs,
we showed that,
despite the clear hardware disadvantage of In-Order (\io) cores vs.\ Out-of-Order (\ooo) equivalent cores,
 online
auto-tuning in \io{} CPUs obtained an average speedup of 1.03 and an
energy efficiency improvement of 39~\% over the SIMD reference in \ooo{} CPUs.


\end{abstract}

\section{Introduction}

High-performance general-purpose embedded processors are evolving with
unprecedented grow in complexity.
ISA (Instruction set architecture)
back-compatibility and energy reduction techniques are among the main
reasons. For sake of software development cost, applications do
not necessarily run in only one target, one binary code may run in processors
from different manufacturers and even in different cores inside a
SoC (System on chip).

Iterative optimization and auto-tuning have been used to automatically find
the best compiler optimizations and algorithm implementations for a given source
code and target CPU.
These tuning approaches have been used to address the
complexity of desktop- and server-class processors (DSCPs). They show
moderate to high performance gains compared to non-iterative compilation,
because default compiler options are usually based on the performance of generic
benchmarks executed in representative hardware.
Usually, such tools need long space exploration times to find
quasi-optimal machine code. Previous work addressed auto-tuning at
run-time~\cite{Voss:2000:AAD:850941.852890,Tiwari:2011:OAC:2058524.2059525,Chen:2012:IOD:2150976.2150983,Ansel:2012:SOA:2380403.2380425}, however previously proposed
auto-tuners are only adapted to applications that run for several minutes
or even hours, such as scientific or data center workload, in order to pay off
the space exploration overhead and overcome the costs of static compilation.

While previous work proposed run-time auto-tuning in DSCP, no work
focused on general-purpose embedded-class processors.
In hand-held devices, applications usually run for a short period of time, imposing a
strong constraint to run-time auto-tuning systems. In
this scenario, a lightweight tool should be employed to explore
pre-identified code optimizations in computing kernels.

We now explain our motivation for developing run-time
auto-tuning tools for general-purpose embedded processors:

     \emph{Embedded core complexity}. The complexity of high-performance
        embedded processors is following the same trend as the complexity of
        DSCP evolved in the last decades. For
        example, current 64-bit embedded-class processors are sufficiently
        complex to be deployed in micro-servers, eventually providing a
        low-power alternative for data center computing. In order to
        address this growing complexity and provide better performance
        portability than static approaches, online auto-tuning is a good
        option.

 \emph{Heterogeneous multi\slash{}manycores}. The power wall is
        affecting embedded systems as they are affecting DSCP, although in a
        smaller scale. Soon, dark silicon may also limit the powerable area in
        embedded SoC. As a consequence, heterogeneous clusters of cores
        coupled to accelerators are one of the solutions being adopted in
        embedded SoC. In the long term, this trend will exacerbate software
        challenges of extracting the achievable computing performance from
        hardware, and run-time approaches may be the only way to improve
        energy efficiency~\cite{Borkar}.

 \emph{ISA-compatible processor diversity}. In the embedded market, a
        basic core design can be implemented by different manufacturers with
        different transistor technologies and also varying configurations.
        Furthermore, customized pipelines may be designed, yet being entirely
        ISA-compatible with basic designs. This diversity of
        ISA-compatible embedded processors facilitates software
        development, however because of differences in pipeline implementations,
        static approaches can only provide sub-optimal performance when
        migrating between platforms.  In addition, contrary to DSCP, in-order (\io{}) cores
        are still a trend in high-performance embedded devices because of
        low-power constraints, and they benefit more from
        target-specific optimizations than out-of-order (\ooo{}) pipelines.

 \emph{Static auto-tuning performance is target-specific}. In
        average, the performance portability of statically auto-tuned code
        is poor when migrating between different
        micro-architectures~\cite{Ansel:2009:PLC:1542476.1542481}. Hence,
        static auto-tuning is usually employed when the execution
        environment is known. On the other hand, the trends of hardware
        virtualization and software packages in general-purpose processors
        result in applications underutilizing the hardware resources, because
        they are compiled to generic micro-architectures. Online
        auto-tuning can provide better performance portability, as previous
        work showed in server-class
        processors~\cite{Ansel:2012:SOA:2380403.2380425}.

 \emph{Ahead-of-time auto-tuning}. In recent Android
        versions (5.0 and later), when an application is installed, native
        machine code is generated from bytecode (ahead-of-time
        compilation). The run-time auto-tuning approach proposed in
        this work could be extended and integrated in such systems to
        auto-tune code to the target core(s) or pre-profile and select
        the best candidates to be evaluated in the run-time phase. Such
        approach would allow auto-tuning to be performed in embedded
        applications with acceptable ahead-of-time compilation
        overhead.

 \emph{Interaction with other dynamic techniques}. Some powerful
        compiler optimizations depend both on input data and the target
        micro-architecture. Constant propagation and loop unrolling are two
        examples. The first can be addressed by dynamically specializing the
        code, while the second is better addressed by an auto-tuning tool.
        When input data and the target micro-architecture are known only at
        program execution, which is usually the case in hand-held devices,
        mixing those two dynamic techniques can provide even
        higher performance improvements. If static versioning is employed, it
        could easily lead to code size explosion, which is not convenient to embedded
        systems. Therefore, run-time code generation and auto-tuning
        is needed.

\medskip

In this article, we propose an online auto-tuning approach for computing
kernels in ARM processors. Existing online auto-tuners regenerate code
using complex compilation chains, which are composed of several transformation
stages to transform source into machine code, leading to important compilation times.
 Our approach consists on
shortening the auto-tuning process, by deploying auto-tuning directly
at the code generation level, through a run-time code generation tool,
called \degoal{}~\cite{Charles2014}.  This allows auto-tuning to be
successfully employed in very short-running kernels, thanks to the low
run-time code generation overhead.
Our very fast online auto-tuner that can quickly explore
the tuning space, and find code variants that are efficient on the running
micro-architecture. The tuning space can have hundreds or even thousands of
valid binary code instances, and hand-held devices may execute applications
that last for a few seconds. Therefore, in this scenario online auto-tuners
have a very strong timing constraint. Our approach addresses this problem with a
two-phase online exploration and by deploying auto-tuning directly at
the level of machine code generation.

The proposed approach is evaluated in a highly CPU-bound (favorable) and
a highly memory-bound (unfavorable) application, to be representative of
all applications between these two extreme conditions. In ARM platforms, the
two benchmarks run during hundreds of milliseconds to only a few seconds. In the
favorable application, the average speedup is 1.26 going up to 1.79,  all
run-time overheads included.

One interesting question that this work tries to answer is if run-time
auto-tuning in simpler and energy-efficient cores can obtain
performance similar to statically compiled code run in more complex and hence
power-hungry cores. The aim is to compare the energy and performance of
\io{} and \ooo{} designs, with the same pipeline and cache configurations,
except for the dynamic scheduling capability. This study would tell us at what
extent run-time auto-tuning of code can replace \ooo{} execution.
However, given that commercial \io{} designs have less resources than \ooo{}
ones (e.g., smaller caches, branch predictor tables), a simulation framework was
employed to perform this experiment. The simulation results show that online
micro-architectural adaption of code to \io{} pipelines can in
average outperform the hand vectorized references run in similar
\ooo{} cores: despite the clear hardware disadvantage, the
proposed approach applied to the CPU-bound application obtained an average
speedup of 1.03 and an energy efficiency improvement of 39~\%.

This article begins by presenting in Section~\ref{sec:auto-tun-motivation} a
motivational example illustrating the potential performance gains if
auto-tuning is pushed to run time. Next,
Section~\ref{sec:auto-tun-approach} presents the problem statement of
auto-tuning short-running kernels at run time and details the proposed
online auto-tuning
approach. The prof of concept is presented through the implementation and
analysis of two case studies in Sections~\ref{sec:auto-tun-expr-setup} and
\ref{sec:auto-tun-results}.
Related work is presented in Section~\ref{sec:related}, and
section~\ref{sec:auto-tun-conclusion} concludes.

\section{Motivational example} \label{sec:auto-tun-motivation}


In this section, we present an experiment supporting the idea that performance achievements could be obtained by the combined use of run-time code specialization and auto-tuning.
The experiment is carried out with a SIMD
 version of the euclidean distance kernel implemented in the Streamcluster
benchmark, manually vectorized in the PARVEC~\cite{Cebrian2014} suite
(originally from the PARSEC 3.0 suite~\cite{bienia11benchmarking}).
In the reference kernel, the dimension of points is a run-time constant,
but given that it is part of the input set, compilers cannot optimize it. In the
following comparisons, we purposefully set the dimension as a compile-time
constant in the reference code to let the compiler (gcc 4.9.3) generate highly
optimized kernels (up to 15~\% of speedup over generic versions). This ensures a
fair comparison with auto-tuned codes.
With \degoal{}~\cite{Charles2014}, in an offline setting, we generated
various kernel versions, by specializing the dimension and auto-tuning the
code implementation for an ARM Cortex-A8 and A9. The auto-tuned parameters mainly
affect loop unrolling and pre-fetching instructions, and are detailed in
Section~\ref{sec:auto-tuning}.
Figure~\ref{fig:motivation} shows the speedups of various generated kernels in
the two core configurations. By analyzing the results, we draw several motivating ideas:

\begin{figure*}
    \centering
    \subfigure[1-D projection of the auto-tuning exploration space. $Dimension=32$\label{fig:motivation-a}]{
	\includegraphics[width=\linewidth]{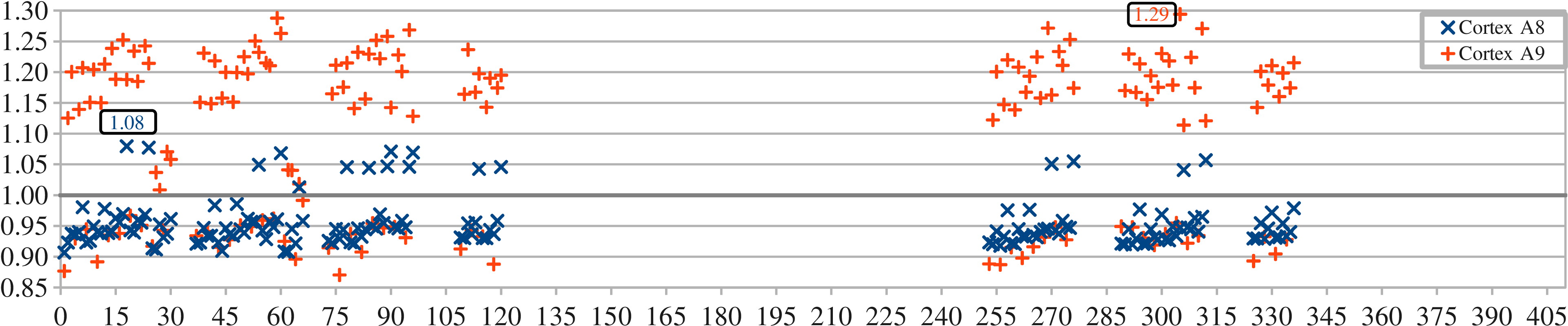}
    }
    \subfigure[1-D projection of the auto-tuning exploration space. $Dimension=128$\label{fig:motivation-b}]{
        \includegraphics[width=\linewidth]{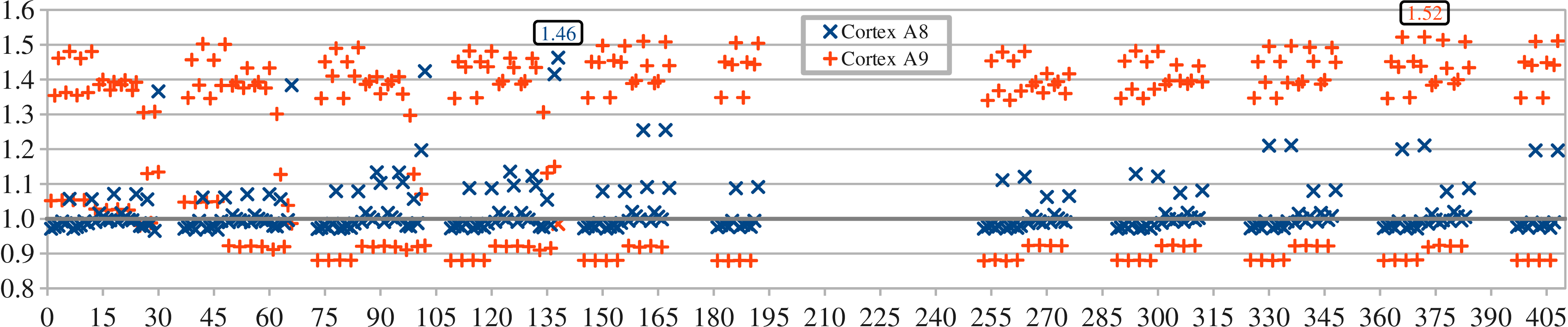}
    }
    \caption{Speedups of euclidean distance kernels statically generated
        with \degoal{}. The reference is a hand vectorized kernel (from PARVEC)
        compiled with gcc 4.9.3 and highly optimized to each target core
        (\texttt{-O3} and \texttt{-mcpu} options). Both \degoal{} and reference
        kernels have the dimension of points specialized (i.e.\ set as a
        compile-time constant). The exploration space goes beyond 600
        configurations, but here it was zoomed in on the main region. The peak
        performance of each core is labeled.  Empty results in the
        exploration space correspond to configurations that could not generate
        code.}
    \label{fig:motivation}
\end{figure*}


 \emph{Code specialization and auto-tuning provide considerable
        speedups even compared to statically specialized and manually vectorized
        code}: Auto-tuned kernel implementations obtained speedups going up to
        1.46 and 1.52 in the Cortex-A8 and A9, respectively.

 \emph{The best set of auto-tuned parameters and optimizations
        varies from one core to another}: In both cases in
        Figure~\ref{fig:motivation}, there is a poor performance portability of
        the best configurations between the two cores. For example, in
        Figure~\ref{fig:motivation-b}, when the best kernel for the
        Cortex-A8 is executed in the A9, the execution time increases by 55~\%,
        compared to the best kernel for the latter. Conversely, the best kernel
        for the A9 when executed in the A8 increases the execution time by 21~\%, compared
        to the most performing kernel.

 \emph{There is no performance correlation between the sets of
        optimizations and input data}: The main auto-tuned parameters are
        related to loop unrolling, which depends on the dimension of points
        (part of the input set). In consequence, the exploration space and the
        optimal solution depend on an input parameter. For example, the configurations
        of the top five peak performances for the A8 in
        Figure~\ref{fig:motivation-b} (configurations 30, 66, 102, 137 and 138)
        have poor performances in Figure~\ref{fig:motivation-a} or simply can
        not generate code with a smaller input set.

The results suggest that, although code specialization and auto-tuning
provide high performance improvements, they should ideally be performed only
when input data and target core are known. In the released input sets for
Streamcluster, the dimensions are 32, 64 and 128, but the benchmark accepts any
integer value.  Therefore, even if the target core(s) was (were) known at
compile time and the code was statically specialized, auto-tuned and
versioned, it could easily lead to code size explosion.

We demonstrate in the following sections that the most important feature of our
approach is that it is fast enough to enable the specialization of run-time
constants combined with online auto-tuning, allowing the generation of
highly optimized code for a target core, whose configuration may not be known prior
compilation.

The optimized kernels shown in this motivational example were statically
auto-tuned. The run-time auto-tuning approach proposed in this
work successfully found optimized kernels whose performance is in average only
6~\% away from the performance of the best kernels statically found (all
run-time overheads included). It is worth observing that the
auto-tuning space has up to 630 valid versions: its exploration took
several hours per dimension and per
platform, even if the benchmark runs for a few seconds.

\section{Online auto-tuning approach} \label{sec:auto-tun-approach}

\begin{figure}
    \centering
    \includegraphics[width=.8\columnwidth]{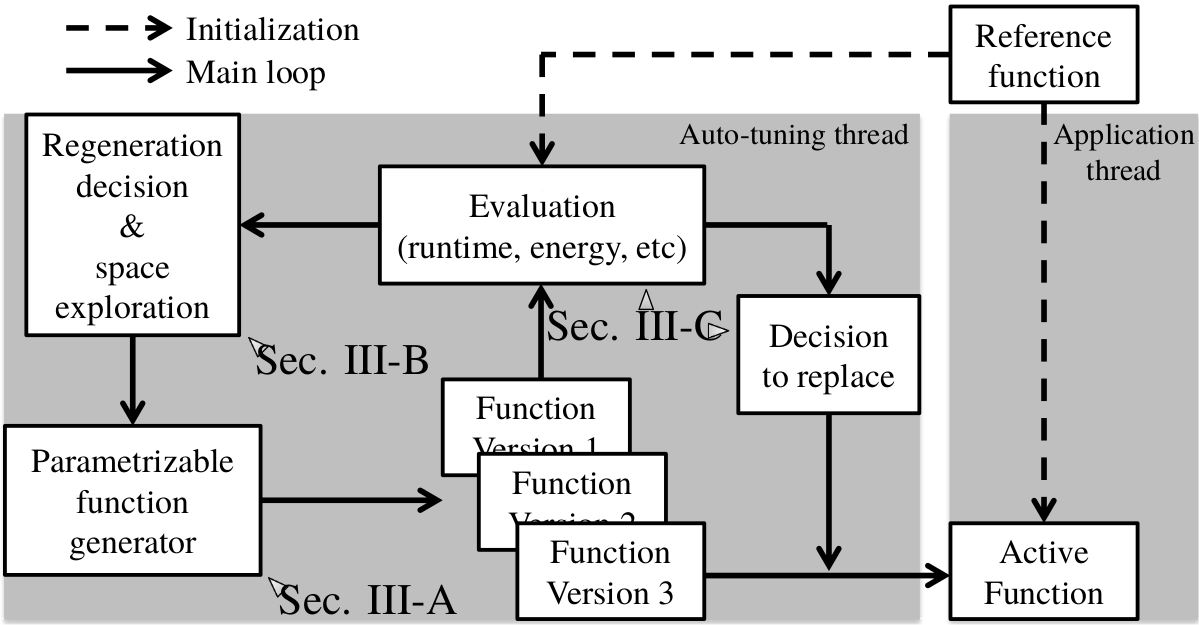}
    \caption{Architecture of the run-time auto-tuning framework.} \label{fig:overview}
\end{figure}

This section describes the approach of the proposed online auto-tuner.
Figure~\ref{fig:overview} presents the architecture of the framework that
auto-tunes a function at run time. At the beginning of the program
execution, a reference function (e.g., C compiled code) is evaluated accordingly
to a defined metric (execution time in the experiments presented here). This reference function
starts as the active function. In parallel to the program execution, the
auto-tuning thread periodically wakes up and decides if it is time to
generate and evaluate a new version. The active
function is replaced by the new one, if its score is better. This approach is applicable to
computing kernels frequently called.

Sections~\ref{sec:auto-tuning}, \ref{sec:regen-n-expl} and \ref{sec:eval-n-repl} describe the
implementation of each block from the main loop of Figure~\ref{fig:overview}.
Section~\ref{sec:prob-statement} presents the problem statement of exploring the tuning space of short-running kernels at run-time.

\subsection{Parametrizable function generator} \label{sec:auto-tuning}

New versions of functions are generated by a tool called \degoal{}. It
implements a domain specific language for run-time code
generation of computing kernels.  It defines a pseudo-assembly
RISC%
-like language, which can be
mixed with standard C code.  The machine code is only generated by \degoal{}
instructions, while the management of variables and code generation decisions
are implemented by \degoal{} pseudo-instructions, optionally mixed with C
code.  The dynamic nature of the language comes from the fact that run-time
information can drive the machine code generation, allowing program and data
specialization.

\degoal{}  supports high-performance ARM processors from the
ARMv7-A architecture, including FP and SIMD instructions.
A set of C functions
were also created to be called in the host application or inside a function
generator to configure the code generation, allowing the evaluation of the
performance impact of various code generation options.

To illustrate some auto-tuning possibilities and how the dynamic language
works, Figure~\ref{fig:auto-tun} presents the \degoal{} code to auto-tune the
euclidean distance implementation, focusing on the main loop of the kernel. This
is the code used in the motivational example presented in
Section~\ref{sec:auto-tun-motivation}, and also in the run-time auto-tuning
experiments later in this work.
Statements between the symbols \mbox{\texttt{\#[}} and
\mbox{\texttt{]\#}} are recognized as the dynamic language, which are
statically converted to standard C code, calling \degoal{} library code to
generate machine instructions at run time.  Other statements are standard C
code. An important symbol in the dynamic language is the sign
\mbox{\texttt{\#()}}: any C expression placed between the parenthesis will
be dynamically evaluated, allowing the inlining of run-time constants or
immediate values.

The kernel generator (called \textit{compilette} in the \degoal{} jargon)
description represented in
Figure~\ref{fig:auto-tun} can generate different machine codes, depending on the
arguments that it receives. In line~1, the first argument is the dimension,
which is specialized in this example. The four following arguments are the
auto-tuned parameters:

\begin{itemize}
    \item \textbf{Hot loop unrolling factor (\mbox{\texttt{hotUF}})}: Unrolls a
        loop and processes each element with a different register, in order to
        avoid pipeline stalls.
    \item \textbf{Cold loop unrolling factor (\mbox{\texttt{coldUF}})}: Unrolls
        a loop by simply copy-pasting a pattern of code, using fewer
        registers, but potentially creating pipeline stalls.
    \item \textbf{Normalized vector length (\mbox{\texttt{vectLen}})}: Defines
        the length of the vector used to process elements in the loop body,
        normalized to the SIMD width when generating SIMD instructions (four in
        the ARM ISA). Longer vectors may benefit code size and speed, because
        instructions that load multiple registers instructions may be generated.
    \item \textbf{Data pre-fetching stride (\mbox{\texttt{pldStride}})}:
        Defines the stride in bytes used in hint instructions to try to
        pre-fetch data of the next loop iteration.
\end{itemize}

Given that the dimension is specialized (run-time constant), we know
exactly how many elements are going to be processed in the main loop. Hence,
between the lines 5 and 21, the pair of \degoal{} instructions \mbox{\texttt{loop}} and
\mbox{\texttt{loopend}} can produce three possible results, depending on the dimension
and the unrolling factors:

\begin{enumerate}
    \item No code for the main loop is generated if the dimension is too small.
        The computation is then performed by a leftover code (not shown in
        Figure~\ref{fig:auto-tun}).
    \item Only the loop body is generated without any branch instruction, if the
        main loop is completely unrolled.
    \item The loop body and a backward branch are generated if more than one
        iteration is needed (i.e.\ the loop is partially unrolled).
\end{enumerate}

The loop body iterates over the coordinates of two points (referenced by
\mbox{\texttt{coord1}} and \mbox{\texttt{coord2}}) to compute the squared
euclidean distance. The computation is performed with vectors, but for the sake of paper conciseness,
variable allocation is not shown in Figure~\ref{fig:auto-tun}. Briefly, in the loop
body, lines 8 and 9 load \mbox{\texttt{vectLen}} coordinates of each point into
vectors, lines 14 and 15 compute the difference, squaring and accumulation, and
finally outside the loop, line 23 accumulates the partial sums in each vector
element of \mbox{\texttt{Vresult}} into \mbox{\texttt{result}}.
Between the lines 6 and 20, the loop body is unrolled by mixing three
auto-tuning effects, whose parameters are highlighted in
Figure~\ref{fig:auto-tun}: the outer \mbox{\texttt{for}} (line~6) simply
replicates \mbox{\texttt{coldUF}} times the code pattern in its body, the inner
\mbox{\texttt{for}} (line~7) unrolls the loop \mbox{\texttt{hotUF}} times by
using different registers to process each pair of coordinates, and finally the
number of elements processed in the inner loop is set through the vector length
\mbox{\texttt{vectLen}}.
In the lines 10 to 13, the last auto-tuned parameter affects a data
pre-fetching instruction: if \mbox{\texttt{pldStride}} is zero, no
pre-fetching instruction is generated, otherwise \degoal{} generates a hint
instruction that tries to pre-fetch the cache line pointed by the address
of the last load plus \mbox{\texttt{pldStride}}.

Besides the auto-tuning possibilities, which are explicitly coded with the
\degoal{} language, a set of C functions can be called to configure code
generation options. In this work, three code optimizations were studied:

\begin{itemize}
    \item \textbf{Instructions scheduling (IS)}: Reorders instructions to avoid
        stall cycles and tries to maximize multi-issues.
    \item \textbf{Stack minimization (SM)}: Only uses FP scratch registers to
        reduce the stack management overhead.
    \item \textbf{Vectorization (VE)}: Generates SIMD instructions to process
        vectors.
\end{itemize}

Most of the explanations presented in this section were given through examples
related to the Streamcluster benchmark, but partial evaluation, loop unrolling
and data pre-fetching are broadly used compiler optimization techniques
that can be employed in almost any kernel-based application.

\begin{figure}
  \centering
  \begin{minipage}[t]{\columnwidth}%
    \lstCplainfileCustom{src/streamcluster4.cdg2}
  \end{minipage}
  \caption{Main loop of the \degoal{} code to auto-tune the euclidean
  distance kernel in the Streamcluster benchmark. The first function parameter
  is the specialized dimension, and the other four are the auto-tuned
  parameters (highlighted variables).}
  \label{fig:auto-tun}
\end{figure}

\subsection{Space exploration: Problem statement}
\label{sec:prob-statement}

The tuning space can be represented by a discrete space with $Nc_{par}$
dimensions, where each dimension corresponds to a parameter being
auto-tuned. Each point in this space correspond to a binary code instance.
To evaluate the tuning space, a
compilette is driven by $Nc_{par} = 7$ parameters $Nc_{i}$ that
control the code generation (\mbox{\texttt{hotUF}}, \mbox{\texttt{coldUF}},
\mbox{\texttt{vectLen}}, \mbox{\texttt{pldStride}}, IS, SM and VE).
More precisely each parameter has a $RangeSize$, in our example in
Figure~\ref{fig:auto-tun},
\mbox{\texttt{vectLen}} belongs to a small set of discrete numbers and
$RangeSize(\mbox{\texttt{vectLen}})=3$, provided by the structure of the kernel.
The number of binary code instances is $N_{codeVariants}$, each one corresponding to a
possible kernel version. Each version is pertinent or not on a
given architecture. We can compute $N_{codeVariants}$ with the following
equation:
\begin{equation}
  N_{codeVariants} = \prod_{i=0}^{Nc_{par}-1}RangeSize(Nc_{i})
\end{equation}

The low-level binary code generator is in charge to adapt a given
variant on a given micro-architecture. A given micro-architecture has
constraints such as number of registers, availability of accelerators and
cache sizes, to name a few. These constraints combined with the structure of the
code explain why the tuning space has holes, where code generation is not
possible as shown in Figure~\ref{fig:motivation}.

The micro-architecture description can also be defined as a
discrete space similar to the tuning space, where each
dimension corresponds to one architectural parameter.
Tables~\ref{tab:sim-params} and \ref{tab:auto-tun-sim-core-abbr} describe
the architecture subspace we
focus on. In this article we focus on a small subset of 2 real and 11 simulated
micro-architectures,
but the market of the embedded systems has many more variants.

This paper proposes a very fast online auto-tuner that can quickly explore
the tuning space, and find code variants that are efficient on the running
micro-architecture. The tuning space can have hundreds or even thousands of
valid binary code instances, and hand-held devices may execute applications
that last for a few seconds. Therefore, in this scenario online auto-tuners
have a very strong timing constraint. Our approach address this problem with a
two-phase online exploration and by deploying auto-tuning directly at
the level of machine code generation.

\subsection{Regeneration decision and space exploration} \label{sec:regen-n-expl}

The regeneration decision takes into account two factors: the regeneration
overhead and the achieved speedup since the beginning of the execution. The
first one allows to
keep the run-time overhead of the tool at acceptable limits if it fails to find
better kernel versions. The second factor acts as an investment, i.e.\
allocating more time to explore the tuning space if previously found solutions
provided sufficient speedups. Both factors are represented as percentage values,
for example limiting the regeneration overhead to 1~\% and investing 10~\% of
gained time to explore new versions.

To estimate the gains, the instrumentation needed in the evaluated functions is
simply a variable that increments each time the function is executed. Knowing
this information and the measured run-time of each kernel, it is possible
to estimate the time gained at any moment. However, given that the reference and
the new versions of kernel have their execution times measured only once, the
estimated gains may not be accurate if the application has phases with very
different behaviors.

Given that the whole space exploration can have hundreds or even thousands of
kernel versions, it was divided in two online phases:

\begin{itemize}
    \item \textbf{First phase}: Explores auto-tuning parameters that have
        an impact on the structure of the code, namely, \mbox{\texttt{hotUF}},
        \mbox{\texttt{coldUF}} and \mbox{\texttt{vectLen}}, but also the
        vectorization option (VE). The previous list is also the order of
        exploration, going from the least switched to the most switched
        parameter. The initial state of the remaining auto-tuning
        parameters
        are determined through pre-profiling.
    \item \textbf{Second phase}: Fixes the best parameters found in the previous
        phase and explores the combinatorial choices of remaining code
        generation options (IS, SM) and \mbox{\texttt{pldStride}}.
\end{itemize}

In our experiments, the range of \mbox{\texttt{hotUF}} and
\mbox{\texttt{vectLen}} were defined by the programmer in a way to avoid running
out of registers, but these tasks can be
automated and dynamically computed by taking into account the code structure
(static) and the available registers (dynamic information). Compared to
\mbox{\texttt{coldUF}}, their ranges are naturally well bounded, providing an
acceptable search space size.

The range of \mbox{\texttt{coldUF}} was limited to 64 after a
pre-profiling phase, because unrolling loops beyond that limit
provided almost no additional speedup.

The last auto-tuned parameter, \mbox{\texttt{pldStride}}, was explored with
the values 32 and 64, which are currently the two possible cache line lengths in
ARM processors.

Finally, to optimize the space exploration, first the tool searches for
kernel implementations that have no leftover code. After exhausting all
possibilities, this condition is softened by gradually allowing leftover
processing.

\subsection{Kernel evaluation and replacement} \label{sec:eval-n-repl}

The auto-tuning thread wakes up regularly to compute the gains and
determine if it is time to regenerate a new function. Each new version is
generated in a dynamically allocated code buffer, and then its performance is
evaluated. When the new code has a better score than that of the active
function, the global function pointer that references the active function is set
to point to the new code buffer.
In order to evaluate a new kernel version, the input data (i.e., processed data)
used in the first and second phases can be either:

\begin{itemize}
    \item \textbf{Real input data only}: Evaluates new kernel versions with real
        data, performing useful work during evaluation, but suffering from
        measurement oscillations between independent runs. These oscillations
        can sometimes lead to wrong kernel replacement decisions.
    \item \textbf{Training \& real input data}: Uses training data with warmed
        caches in the first phase and real data in the second one. A training
        input set with warmed caches results in very stable measurements, which
        ensure good choices for the active function. Since no useful work is
        performed, using training data is only adapted to kernels that are
        called sufficient times to consider the overhead of this technique
        negligible, and to kernels that have no side effect.
        In the second phase, the usage of real data is mandatory, because the
        adequacy of pre-fetching instruction depends on the interaction of
        the real data and code with the target pipeline.
\end{itemize}

When the evaluation uses real data, the performance of the kernel is
obtained by simply averaging the run-times of a pre-determined number of
runs.

When the kernel uses a training input data, the measurements are filtered. We
took the worst value between the three best values of groups with five
measurements. This technique
filters unwanted oscillations caused by hardware (fluctuations in the pipeline,
caches and performance counters) and software (interruptions). In the studied
platforms, stable measurements were observed, with virtually no oscillation
between independent runs (in a Cortex-A9, we measured oscillations of less
than 1~\%).

The decision to replace the active function by a new version is taken by simply
comparing the calculated run-times.

\section{Experimental setup} \label{sec:auto-tun-expr-setup}

This section presents the experimental setup. First, we detail the hardware and
simulation platforms employed in the experiments. Then, the chosen applications
for two case studies are described.
The kernel run-times and the auto-tuning overhead are measured through
performance counters.

\subsection{Hardware platforms}

Two ARM boards were used in the experiments.  One is the Snowball equipped
with a dual Cortex-A9 processor~\cite{CalaoSystems2011}, an \ooo{}
pipeline. The board runs the Linaro 11.11 distribution with a Linux 3.0.0
kernel. The other is the BeagleBoard-xM, which has an \io{} Cortex-A8
core~\cite{BeagleBoard.org2010}. The board runs a Linux 3.9.11 kernel with an
Ubuntu 11.04 distribution.

In order to evaluate the online auto\hyp{}tuning overhead per core, the
benchmarks are forced to execute in one core through the Linux command
\mbox{\texttt{taskset}}, because the code regeneration is performed in a
separated thread.
The fact that auto\hyp{}tuning in performed in single\hyp{}cores is not a
limitation of the framework. The proposed approach can be extended to locally
auto\hyp{}tune the code of a kernel in each core of a heterogeneous multicore
system.

\subsection{Simulation platform}

A micro\hyp{}architectural simulation framework~\cite{Endo},
\cite{Endo:2015:MSE:2693433.2693440} was used to simulate 11 different core
configurations. It is a modified version of the
gem5~\cite{Binkert:2011:GS:2024716.2024718} and
McPAT~\cite{Li:2013:MFM:2445572.2445577} frameworks, for performance and
power\slash{}area estimations, respectively. Table~\ref{tab:sim-params} shows the
main configurations of the simulated cores. The 11 configurations were obtained
by varying the pipeline type (\io{} and \ooo{} cores) and the number of
VPUs (FP\slash{}SIMD units) of one-, two- and three\hyp{}way basic pipelines. The
single\hyp{}issue core is \io{} an has only one VPU.
Dual\hyp{}issue cores can have one or two VPUs, while
triple\hyp{}issue cores can have one, two or three VPUs.
In McPAT, the temperature is fixed at 47~\degree{}C and the transistor
technology is 28~nm.
Table~\ref{tab:auto-tun-sim-core-abbr} shows the abbreviations used to identify
each core design and their CPU areas.

\begin{sidewaystable}
\begin{center}
\caption{Main parameters of the simulated cores.}
\label{tab:sim-params}
\begin{threeparttable}[b]
\begin{tabular}{|ll|c|c|c|} \hline
\multicolumn{2}{|c|}{Parameter}                                                      & Single-issue            & Dual-issue              & Triple-issue     \\ \hline
\multicolumn{2}{|l|}{Pipeline type}                                                  & IO only           & IO or OOO& IO or OOO \\ \hline
\multicolumn{2}{|l|}{Core clock}                                                     & 1.4~GHz                 & 1.6~GHz                 & 2.0~GHz          \\ \hline
\monocol{|l|}{DRAM}                         & Size/clock/latency (ns)            & 256~MB/933~MHz/81   & 256~MB/933~MHz/81   & 256~MB/933~MHz/81  \\ \hline
\monocol{|l|}{L2}                           & Size/assoc./lat./MSHRs/WBs     & 512~kB/8/3/8/16 & 1024~kB/8/5/8/16& 2048~kB/16/8/11/16 \\ \hline
\monocol{|l|}{L1-I}                         & Size/assoc./lat./MSHRs           & 32~kB/2/1/2       & 32~kB/2/1/2       & 32~kB/2/1/2      \\ \hline
\monocol{|l|}{L1-D}                         & Size/assoc./lat./MSHRs/WBs     & 32~kB/4/1/4/4   & 32~kB/4/1/5/8   & 32~kB/2/1/6/16   \\ \hline
\monocol{|l|}{Stride prefet.}               & Cache level/degree/buffer size     & 1/1/8               & 1/1/12              & 2/1/16                \\ \hline
\monocol{|l|}{\multirow{2}{*}{Branch pred.}}& Global/local history entries (bits)  & 256 (2)/-           & 4096 (2)/-          & 4096 (2)/1024 (3)  \\ \cline{2-5}
\monocol{|c|}{}                             & BTB/RAS entries                      & 256/8                 & 4096/16               & 4096/48        \\ \hline
\multicolumn{2}{|l|}{Front\hyp{}end/back\hyp{}end width}                                     & 1/1                   & 2/4                   & 3/7              \\ \hline
\multicolumn{2}{|l|}{INT/FP pipeline depth (+ extra OOO stages)}          & 8/10                  & 8/12 (+3)             & 9/18 (+6)            \\ \hline
\multicolumn{2}{|l|}{Physical INT/FP registers\tnote{2}}            & -                     & 82/256                & 90/256           \\ \hline
\multicolumn{2}{|l|}{ITLB/DTLB/IQ/LSQ/ROB\tnote{2}\,\, entries}                    & 32/32/16/8~each/-     & 64/64/32/12~each/40   & 128/128/48/16~each/60 \\ \hline
\monocol{|l|}{\multirow{2}{*}{INT units}}   & ALU/MUL execution ports              & 1/1                   & 2/1                   & 2/1             \\ \cline{2-5}
\monocol{|c|}{}                             & ADD/MUL cycles                       & 1/4                   & 1/4                   & 1/4             \\ \hline
\monocol{|l|}{\multirow{2}{*}{FP/SIMD}}     & Execution ports                        & 1                       & 1 or 2                  & 1, 2 or 3         \\ \cline{2-5}
\monocol{|c|}{}                             & VADD/VMUL/VMLA cycles              & 3/4/6               & 4/5/8               & 10/12/20      \\ \hline
\monocol{|l|}{\multirow{2}{*}{Load/store}}  & Execution ports                        & 1 shared                & 1 shared                & 1 for each        \\ \cline{2-5}
\monocol{|c|}{}                             & Load/store cycles                    & 1/1                   & 2/1                   & 3/2             \\ \hline
\end{tabular}
\begin{tablenotes}
    \item [1] Over-dimensioned to compensate the lack of L2-TLB.
    \item [2] For \ooo{} only.
\end{tablenotes}
\end{threeparttable}
\end{center}
\end{sidewaystable}

\begin{table}
\begin{center}
  \caption{Abbreviation of the simulated core designs and CPU areas}
  \label{tab:auto-tun-sim-core-abbr}
  \begin{tabular}{|l|c|c|c|c|c|c|} 
	\hline
	\monocol{|c|}{\multirow{2}{*}{Abbrev.}} 
	& \multirow{2}{*}{Width} 
	& \multirow{2}{*}{Type} 
	& \multirow{2}{*}{VPUs} 
	& \multicolumn{3}{c|}{Area (\SI{}{\square\milli\meter})}  \\ \cline{5-7}
    & & & & Core & L2 & Total \\ \hline \hline
    SI-I1 & 1 & IO  & 1 & 0.45 & 1.52 & 1.97 \\ \hline
	 TI-I1 & 3 & IO  & 1 & 1.81 & 5.88 & 7.70 \\ \hline
 TI-I2 & 3 & IO  & 2 & 2.89 & 5.88 & 8.78 \\ \hline
    DI-I1 & 2 & IO  & 1 & 1.00 & 3.19 & 4.19 \\  \hline
	TI-I3 & 3 & IO  & 3 & 3.98 & 5.88 & 9.86 \\ \hline
    DI-I2 & 2 & IO  & 2 & 1.48 & 3.19 & 4.67 \\ \hline
 TI-O1 & 3 & OOO & 1 & 2.08 & 5.88 & 7.97 \\ \hline
    DI-O1 & 2 & OOO & 1 & 1.15 & 3.19 & 4.34 \\ \hline
 TI-O2 & 3 & OOO & 2 & 3.21 & 5.88 & 9.10 \\ \hline
    DI-O2 & 2 & OOO & 2 & 1.67 & 3.19 & 4.86 \\ \hline
 TI-O3 & 3 & OOO & 3 & 4.35 & 5.88 & 10.2 \\ \hline
  \end{tabular}
\end{center}
\end{table}

\subsection{Benchmarks}

Two kernel-based applications were chosen as case studies to evaluate the
proposed online auto-tuning approach. To be representative, one benchmark
is CPU-bound and the other is memory-bound.
In both applications, the evaluated kernels correspond to more than 80~\% of
execution time.
The benchmarks were compiled with gcc 4.9.3 (gcc 4.5.2 for Streamcluster
binaries used in the simulations) and the default PARSEC flags (\texttt{-O3
-fprefetch-loop-arrays} among others). The NEON flag
(\texttt{-mfpu=neon}) is set to allow all 32 FP registers to be used. The target
core is set (\texttt{-mcpu} option) for the real platforms and the ARMv7-A
architecture (\texttt{-march=armv7-a}) for binaries used in the simulations. The
\degoal{} library was also compiled for the ARMv7-A architecture, which covers
all real and simulated CPUs.

The first kernel is the euclidean distance computation in the Streamcluster
benchmark from the PARSEC 3.0 suite. It solves the online clustering problem.
Given points in a space, it tries to assign them to nearest centers. The
clustering quality is measured by the sum of squared distances. With high space
dimensions, this benchmark is CPU-bound~\cite{bienia11benchmarking}. In
the compilette definition, the dimension (run-time constant) is specialized.
The \mbox{\texttt{simsmall}} input set is evaluated with the dimensions 32
(original), 64 and 128 (as in the native input set), which are referred as
small, medium and large input sets, respectively.

The second kernel is from VIPS, an image processing application. A linear
transformation is applied to an image with the following command line:

\begin{verbatim}
vips im_lintra_vec MUL_VEC input.v ADD_VEC output.v
\end{verbatim}

Here, \mbox{\texttt{input.v}} and \mbox{\texttt{output.v}} are images in the
VIPS XYZ format, and \mbox{\texttt{MUL\_VEC}}, \mbox{\texttt{ADD\_VEC}} are
respectively FP vectors of the multiplication and addition factors for each band
applied to each pixels in the input image. Given that pixels are loaded and
processed only once, it is highly memory-bound. Indeed, the auto-tuned
parameters explored in this work are not suitable for a memory-bound
kernel. However, this kind of kernel was also evaluated to cover unfavorable
situations and show that negligible overheads are obtained. In the compilette description,
two run-time constants, the number of bands and the width of the image, are
specialized. Three input sets were tested: \mbox{\texttt{simsmall}} ($1600$ x
$1200$), \mbox{\texttt{simmedium}} ($2336$ x $2336$) and
\mbox{\texttt{simlarge}} ($2662$ x $5500$).

\subsection{Evaluation methodology}

The execution time of the applications is measured through the Linux command
\mbox{\texttt{time}}. Between 3 and 20 measurements were collected depending on
observed oscillations.

The proposed online approach can auto\hyp{}tune both SISD\footnote{Single
instruction, single data.} and SIMD codes during the
application execution. In order to allow a fair comparison between the proposed approach
and the references, the auto\hyp{}tuning internally generates and evaluates
both SIMD and SISD code, but when comparing to the SISD reference, SIMD
generated codes are ignored and only SISD kernels can be active in the
application, and vice\hyp{}versa. In a real scenario, the performance achieved
by the proposed approach is the best among the SISD and SIMD results presented in
this work. The SISD reference is taken as initial active function, because
this is a realistic scenario.

In the real platforms, the tuning space is also statically explored to find the
best kernel implementation per platform and per input set. In order to limit
prohibitive exploration times, the search was limited to only optimal solutions
(no leftovers) in Streamcluster and to at least 1000 points in the search space
for VIPS, because some input sets had only few optimal solutions. Part of this
exploration is shown in Figure~\ref{fig:motivation}.


\section{Experimental results} \label{sec:auto-tun-results}

This section presents the experimental results of the proposed online
auto-tuning approach in a CPU- and a memory-bound kernels.
First, the results obtained in real and simulated platforms are presented.
Then, experiments with varying workload size is discussed.
Finally, an analysis of auto-tuning parameter correlation to pipeline designs is presented.

\subsection{Real platforms}

Table~\ref{tab:hwres-all-runtimes} presents the execution times of all
configurations studied of the two benchmarks in the real platforms.

\begin{figure}
\centering
\subfigure[Streamcluster in Cortex-A8 \label{fig:hwres-stream-all-a}]{%
	\includegraphics[width=.48\columnwidth]{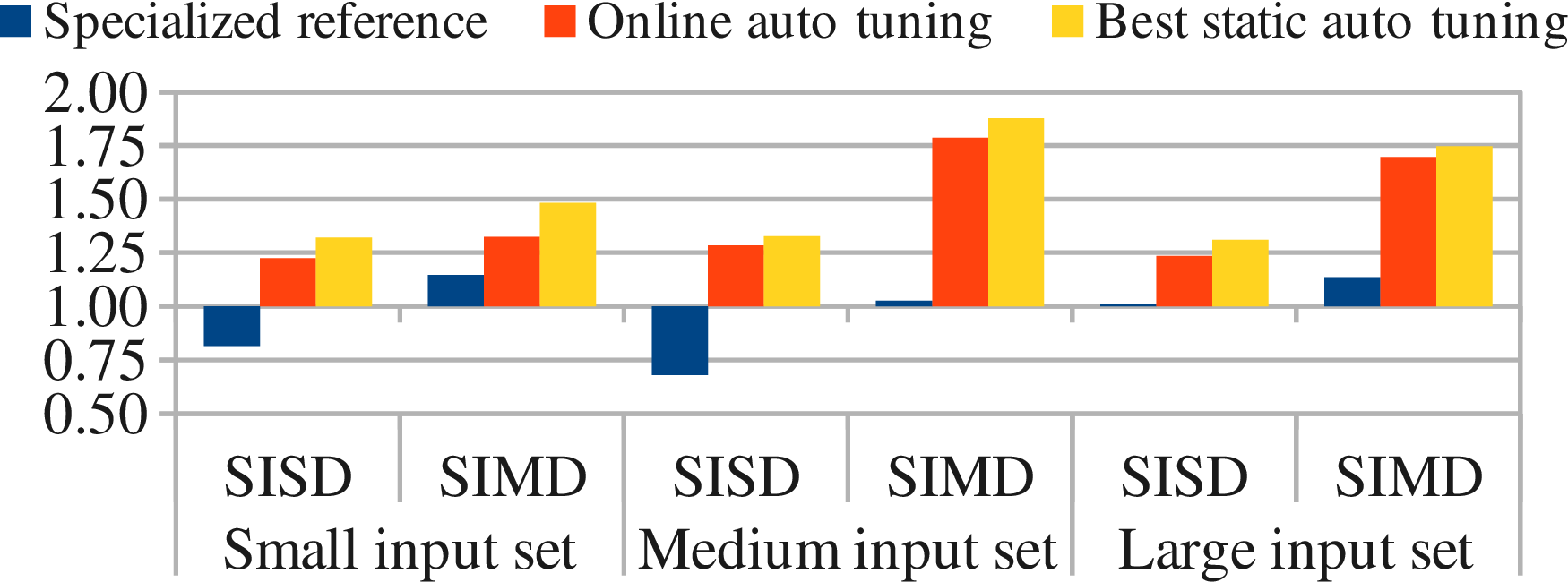}
	}
\subfigure[Streamcluster in Cortex-A9 \label{fig:hwres-stream-all-b}]{%
	\includegraphics[width=.48\columnwidth]{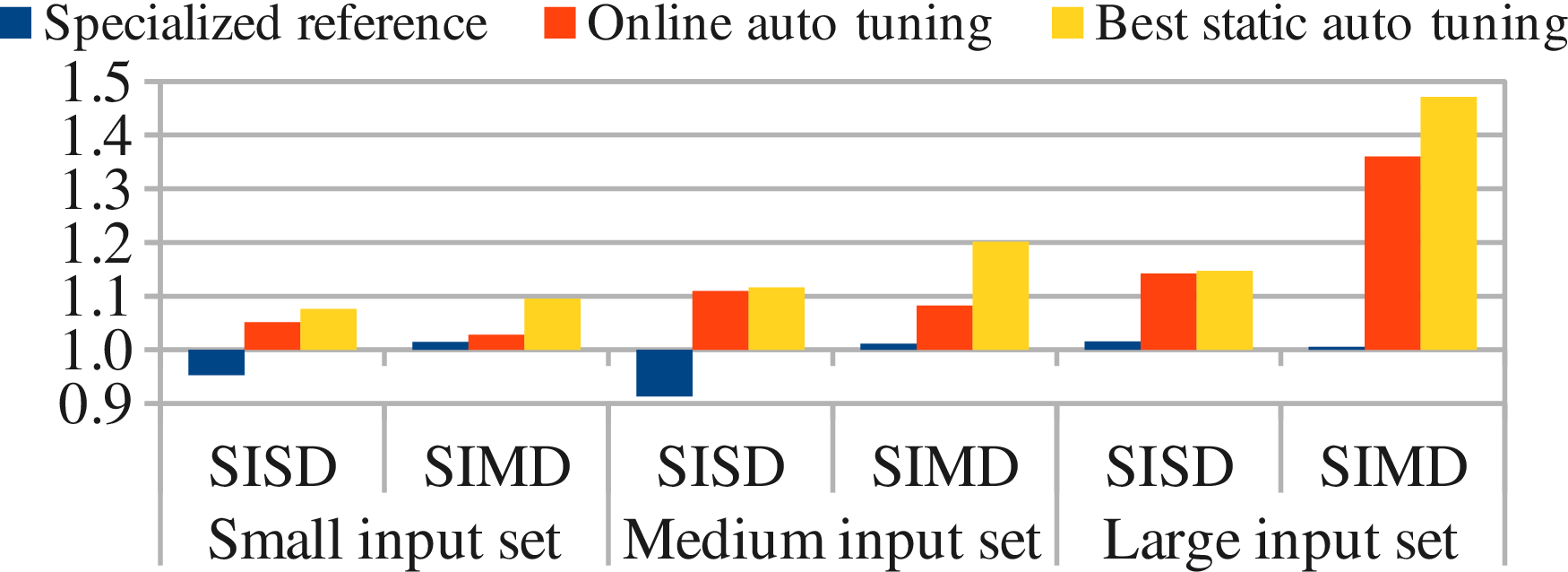}
	}

\subfigure[VIPS in Cortex-A8 \label{fig:hwres-stream-all-c}]{%
	\includegraphics[width=.48\columnwidth]{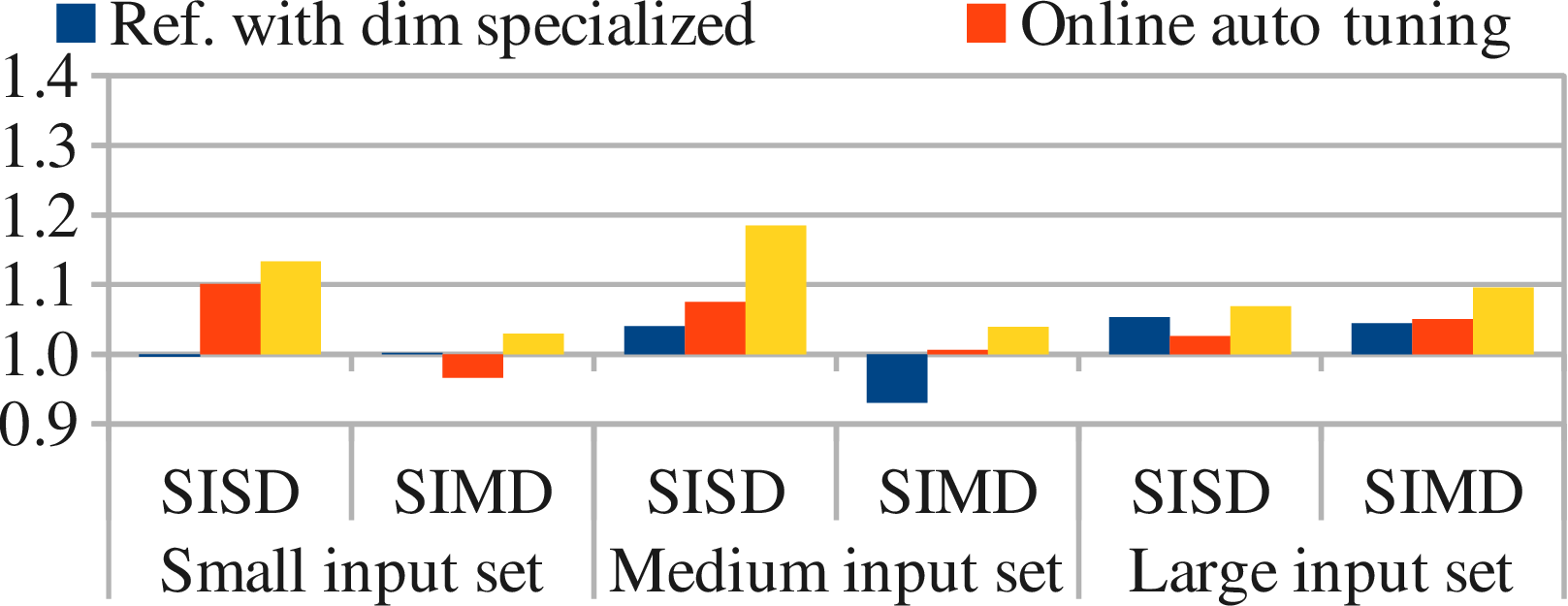}
	}
\subfigure[VIPS in Cortex-A9 \label{fig:hwres-stream-all-d}]{%
	\includegraphics[width=.48\columnwidth]{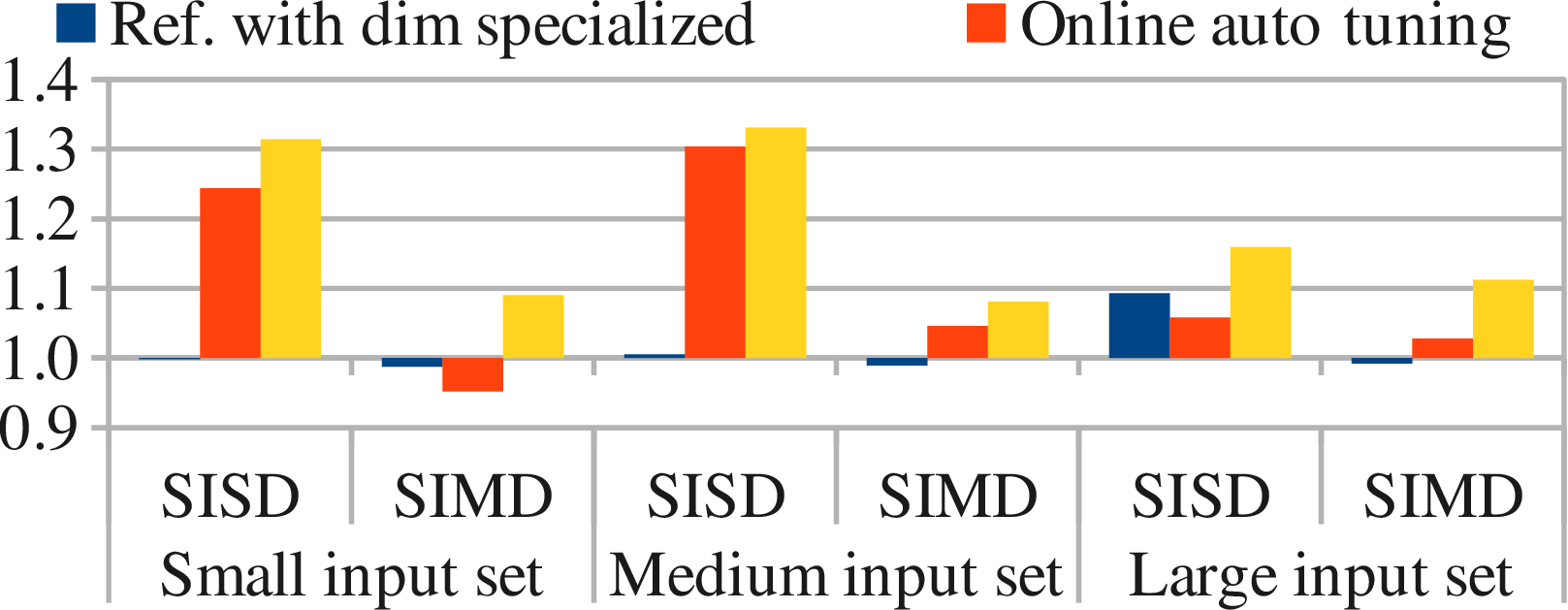}
	}
    \caption{Speedup of the specialized reference and the auto-tuned
    applications in the real platforms (normalized to the reference benchmarks).}
\end{figure}

Figures~\ref{fig:hwres-stream-all-a} and \ref{fig:hwres-stream-all-b} show the
speedups obtained in the Streamcluster
benchmark.  Run-time auto-tuning provides
average speedup factors of 1.12 in the Cortex-A8 and 1.41 in the A9. The speedup sources
come mostly from micro-architectural adaption, because even if the reference
kernels are statically specialized, they can not provide significant speedups.
The online auto-tuning performance is only 4.6~\% and 5.8~\% away from those
of the best statically auto-tuned versions, respectively for the
A8 and A9.

\begin{sidewaystable}
\begin{center}
\begin{threeparttable}[b]
  \caption{Execution time (seconds) of the benchmarks with the original and
  specialized reference kernels, and with the online auto\hyp{}tuned and the
  best statically auto\hyp{}tuned kernels, in the real platforms (all
  run\hyp{}time overheads included).}
  \label{tab:hwres-all-runtimes}
  \begin{tabular}{|l|l|l|c|c|c|c|c|c|c|c|} \hline
    \monocol{|c|}{\multirow{2}{*}{Benchmark}} & \monocol{c|}{\multirow{2}{*}{Input}} & \multirow{2}{*}{Version} & \multicolumn{4}{c|}{Cortex-A8} & \multicolumn{4}{c|}{Cortex-A9} \\ \cline{4-11}
                                  &                       &      & Ref. & Spec.\ Ref.&  O-AT  & BS-AT & Ref. & Spec.\ Ref.&  O-AT  & BS-AT  \\ \hline
    \multirow{6}{*}{Streamcluster}&\multirow{2}{*}{Small} & SISD & 9.75 & 10.2       & 9.26   & 9.06  & 3.26 & 4.00       & 2.66   & 2.47   \\ \cline{3-11}
                                  &                       & SIMD & 3.84 & 3.79       & 3.74   & 3.51  & 3.33 & 2.90       & 2.51   & 2.24   \\ \cline{2-11}
                                  &\multirow{2}{*}{Medium}& SISD & 19.9 & 21.8       & 17.9   & 17.8  & 7.54 & 11.1       & 5.87   & 5.68   \\ \cline{3-11}
                                  &                       & SIMD & 7.13 & 7.05       & 6.59   & 5.93  & 9.09 & 8.86       & 5.09   & 4.84   \\ \cline{2-11}
                                  &\multirow{2}{*}{Large} & SISD & 46.8 & 46.1       & 41.0   & 40.8  & 14.8 & 14.7       & 12.0   & 11.3   \\ \cline{3-11}
                                  &                       & SIMD & 15.1 & 15.0       & 11.1   & 10.2  & 17.2 & 15.1       & 10.1   & 9.84   \\ \hline
    \multirow{6}{*}{VIPS lintra}  &\multirow{2}{*}{Small} & SISD & 0.841& 0.842      & 0.676  & 0.640 & 0.502& 0.504      & 0.456  & 0.443  \\ \cline{3-11}
                                  &                       & SIMD & 0.556& 0.563      & 0.584  & 0.510 & 0.455& 0.454      & 0.471  & 0.442  \\ \cline{2-11}
                                  &\multirow{2}{*}{Medium}& SISD & 2.30 & 2.29       & 1.76   & 1.73  & 1.47 & 1.41       & 1.37   & 1.24   \\ \cline{3-11}
                                  &                       & SIMD & 1.47 & 1.48       & 1.40   & 1.36  & 1.31 & 1.41       & 1.31   & 1.26   \\ \cline{2-11}
                                  &\multirow{2}{*}{Large} & SISD & 26.6 & 24.3       & 25.1   & 22.9  & 10.1 & 9.63       & 9.88   & 9.49   \\ \cline{3-11}
                                  &                       & SIMD & 24.7 & 24.9       & 24.0   & 22.2  & 10.4 & 10.0       & 9.94   & 9.54   \\ \hline
  \end{tabular}
  \begin{tablenotes}
    \item [] Spec.\ ref.: Reference kernel specialized as in the auto\hyp{}tuned
    versions.
    \item [] O-AT: Online auto\hyp{}tuned kernel.
    \item [] BS-AT: Best statically auto\hyp{}tuned kernel.
  \end{tablenotes}
\end{threeparttable}
\end{center}

\begin{center}
\begin{threeparttable}[b]
  \caption{Statistics of online auto\hyp{}tuning in the Cortex\hyp{}A8 and A9
(SISD / SIMD separated, or average if minor variations).}
  \label{tab:online-stats}
  \begin{tabular}{|l|l|c|c|c|c|c|c|c|c|c|} \hline
      \multirow{3}{*}{Bench.}  &\multirow{3}{*}{Input set}&\multirow{3}{*}{\parbox{1.4cm}{\centering{}Explo-rable versions}}& \multirow{3}{*}{\parbox{1.8cm}{\centering{}Exploration limit in one run}}&\multicolumn{7}{c|}{Run\hyp{}time regeneration and space exploration} \\ \cline{5-11}
                                  &                          &       &               &\multirow{2}{*}{\parbox{1cm}{\centering{}Kernel calls}} &\multicolumn{2}{c|}{Explored} &\multicolumn{2}{c|}{Overhead to bench. run-time} &\multicolumn{2}{c|}{Duration to kernel life} \\ \cline{6-11}
                                  &                          &       &               &                        &A8 &A9 &A8            &A9             &A8           &A9     \\ \hline
    \multirow{3}{*}{\parbox{1cm}{Stream-cluster}}&Small      &390    &43-49          &\multirow{3}{*}{5315388}&49 &49 &0.2 \% (11 ms)&0.4 \% (9.2 ms)&13 / 4.4~\%  &32~\%  \\
                                  &Medium                    &510    &55-61          &                        &58 &61 &0.2 \% (17 ms)&0.3 \% (15 ms) &6.3 / 2.7~\% &22~\%  \\
                                  &Large                     &630    &67-73          &                        &67 &73 &0.2 \% (30 ms)&0.2 \% (26 ms) &5.6 / 1.8~\% &15~\%  \\ \hline
    \multirow{3}{*}{VIPS}         &Small                     &858    &106-112        &1200                    &44 &28 &4.2 \% (26 ms)&2.5 \% (12 ms) &100~\%       &100~\% \\
                                  &Medium                    &330    &39-45          &2336                    &40 &42 &0.9 \% (14 ms)&1.0 \% (14 ms) &18~\%        &66~\%  \\
                                  &Large                     &596    &73-79          &5500                    &75 &71 &0.3 \% (71 ms)&0.8 \% (78 ms) &28~\%        &86~\%  \\ \hline
    \end{tabular}
\end{threeparttable}
\end{center}
\end{sidewaystable}

Still in Streamcluster, from Table~\ref{tab:hwres-all-runtimes} we can also note
that in the A9, the SIMD reference versions are in average
11~\% slower than the SISD references, because differently from the SISD
reference data pre\hyp{}fetching instructions are not generated by gcc in the
SIMD code. The online approach can effectively take advantage of SIMD
instructions in the A9, providing in average a speedup of 1.41 compared to the
SISD reference code and 1.13 compared to
dynamically auto\hyp{}tuned SISD code.

Figures~\ref{fig:hwres-stream-all-c} and \ref{fig:hwres-stream-all-d} show the
speedups obtained in the VIPS application. Even with the hardware bottleneck being
the memory hierarchy, in average the proposed approach can still speed up the execution
by factors of 1.10 and 1.04 in the A8 and A9, respectively. Most of the speedups
come from SISD versions (SIMD performances almost matched the
references), mainly because in the reference code
run-time constants are reloaded in
each loop iteration, differently from the compilette implementation. In average, online
auto-tuning performances are only 6~\% away from the best static ones.

Table~\ref{tab:online-stats} presents the auto-tuning statistics in both
platforms. For each benchmark and input set, it shows that between 330 and 858
different kernel configurations could be generated, but in one run this space is
limited between 39 and 112 versions, thanks to the proposed two phase
exploration (Section~\ref{sec:regen-n-expl}). The online statistics gathered in
the experiments are also presented. In most cases, the exploration ends very
quickly, specially in Streamcluster, in part because of the investment factor.
Only with the small input in VIPS, the auto-tuning did not end during its
execution, because it has a large tuning space and VIPS executes
during less than 700~ms. The overhead of the run-time approach is
negligible, between only 0.2 and 4.2~\% of the application run-times were spent
to generate and evaluate from 28 to 75 kernel versions.

\subsection{Simulated cores}

\begin{figure}
    \centering
    \subfigure[Small input set.]{\includegraphics[width=\columnwidth]{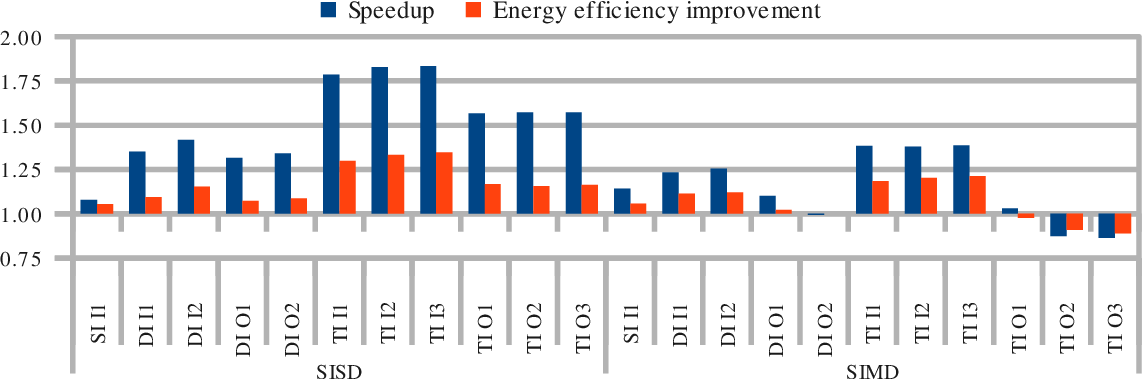}}
    \subfigure[Medium input set.]{\includegraphics[width=\columnwidth]{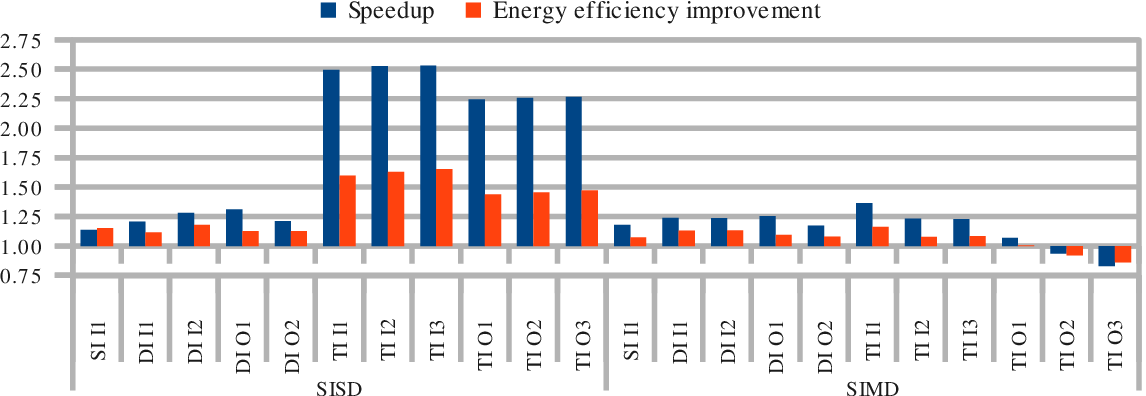}}
    \subfigure[Large input set.]{\includegraphics[width=\columnwidth]{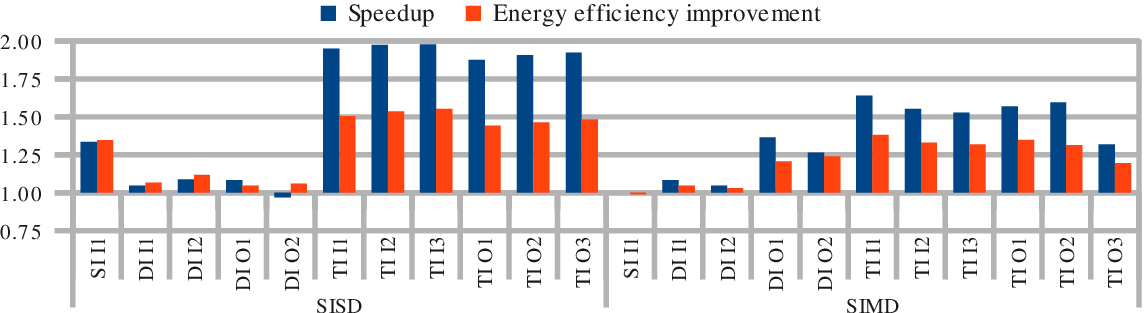}}
    \caption{%
	Speedup and energy efficiency improvement of online  auto-tuning over the references codes in the Streamcluster benchmark,  simulating the 11 cores. Core abbreviations are listed in     Table~\ref{tab:auto-tun-sim-core-abbr}.
}
    \label{fig:simres}
\end{figure}

Figure~\ref{fig:simres} shows the simulated energy and performance of the
reference and online auto-tuning versions of the Streamcluster benchmark.
In the SISD comparisons, run-time auto-tuning can find kernel
implementations with more
ILP (Instruction-level parallelism) than the reference code, specially
remarkable in the long triple-issue pipelines. The average speedup is 1.58.
In the SIMD comparisons, the reference kernel naturally benefits from the
parallelism of vectorized code, nonetheless online auto-tuning can provide
an average speedup of
1.20. Only 6 of 66 simulations showed worse performance, mostly in big cores
that quickly executed the benchmark.

\begin{figure}
    \centering

    \subfigure[Speedup of \io{} vs \ooo{} cores.\label{fig:auto-tun-simres-io-ooo-perf}]{\includegraphics[width=0.49\textwidth]{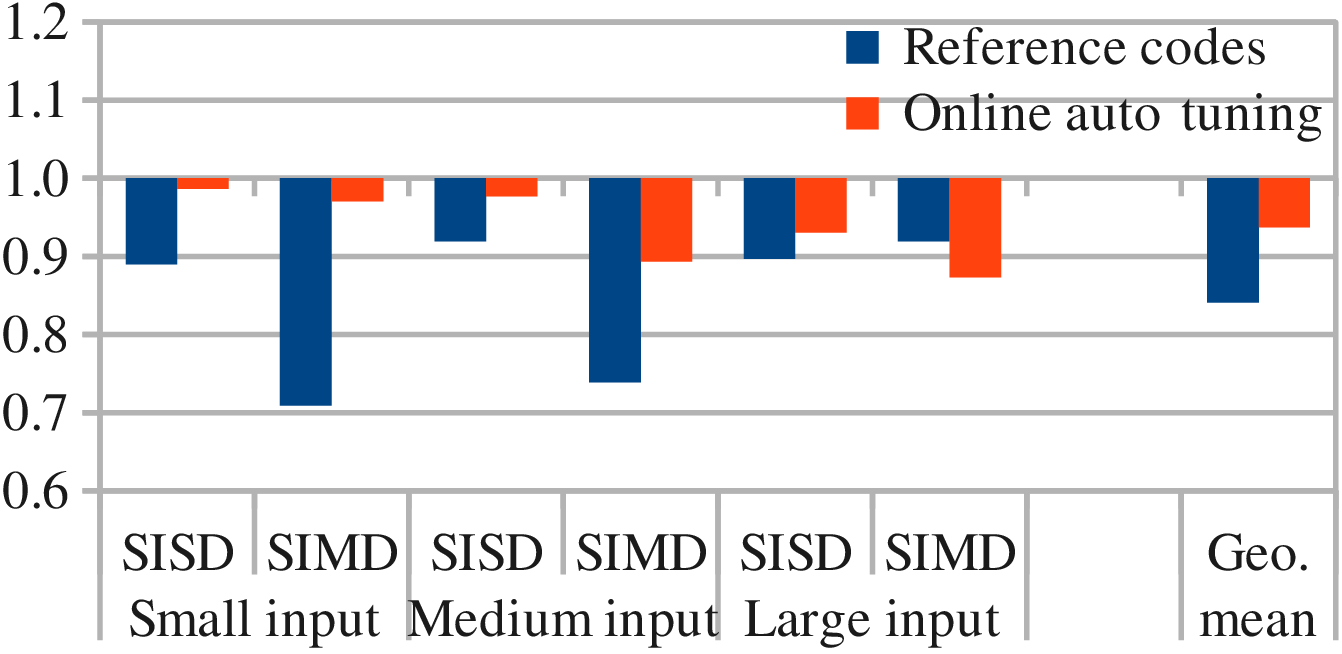}}
\hspace*{\fill}
    \subfigure[Energy efficiency of \io{} vs \ooo{} cores.\label{fig:auto-tun-simres-io-ooo-energy}]{\includegraphics[width=0.49\textwidth]{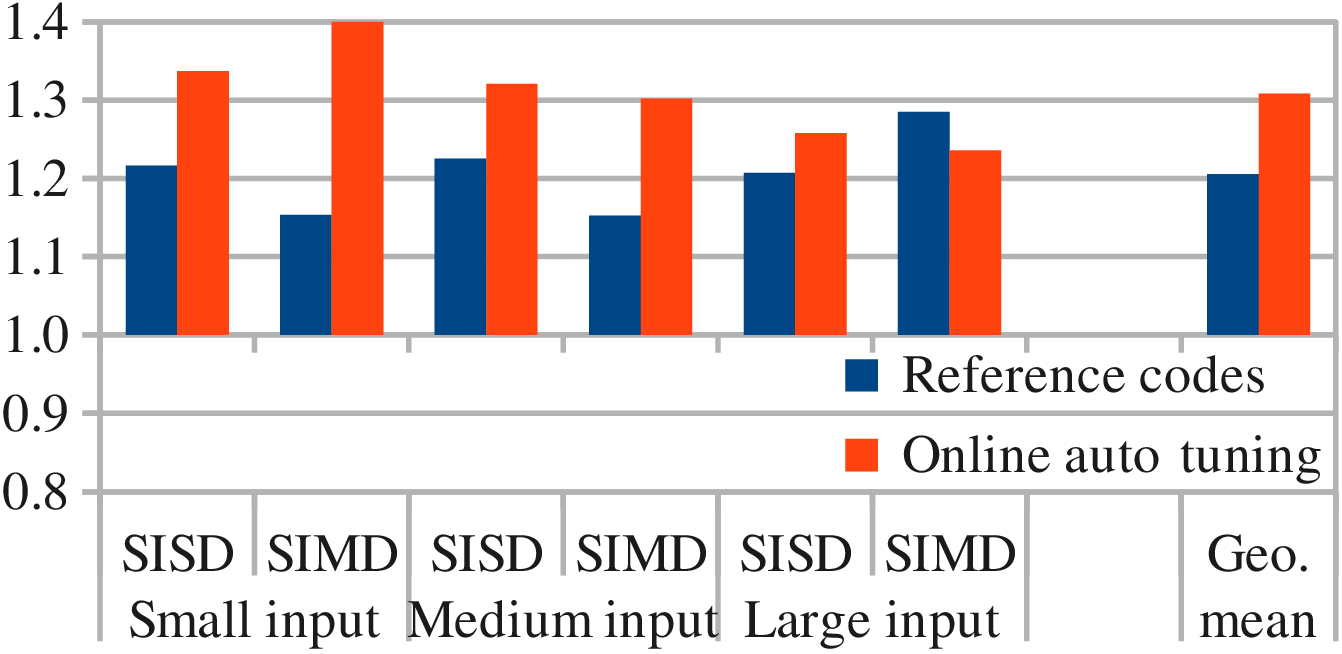}}

    \subfigure[Energy efficiency and performance improvement of online auto\hyp{}tuning in \io{} vs reference in \ooo{} cores.\label{fig:auto-tun-simres-io-at-ooo-ref}]{\includegraphics[width=0.55\textwidth]{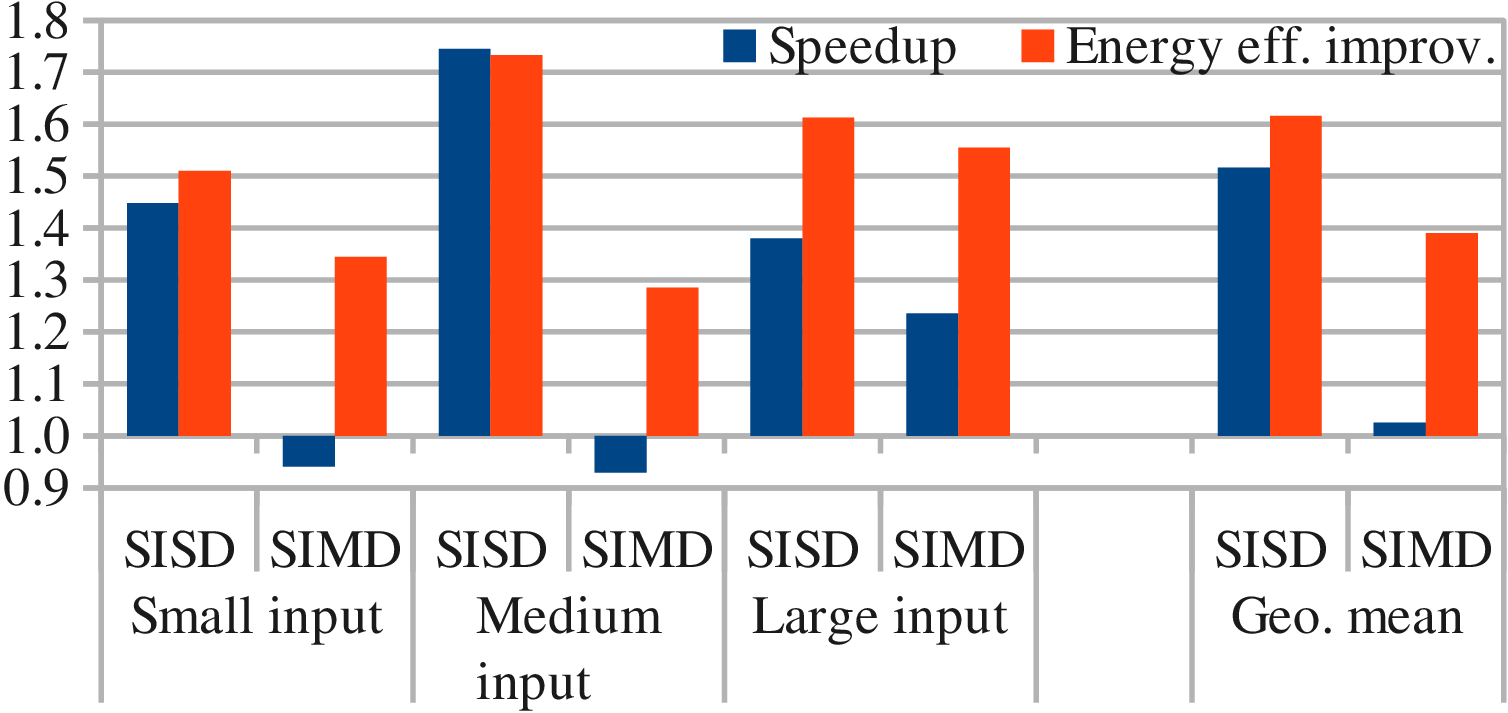}}
\hspace*{\fill}
    \subfigure[Area overhead of \ooo{} vs \io{} (excluding L2 caches).\label{fig:auto-tun-simres-io-ooo-area}]{\includegraphics[width=0.35\textwidth]{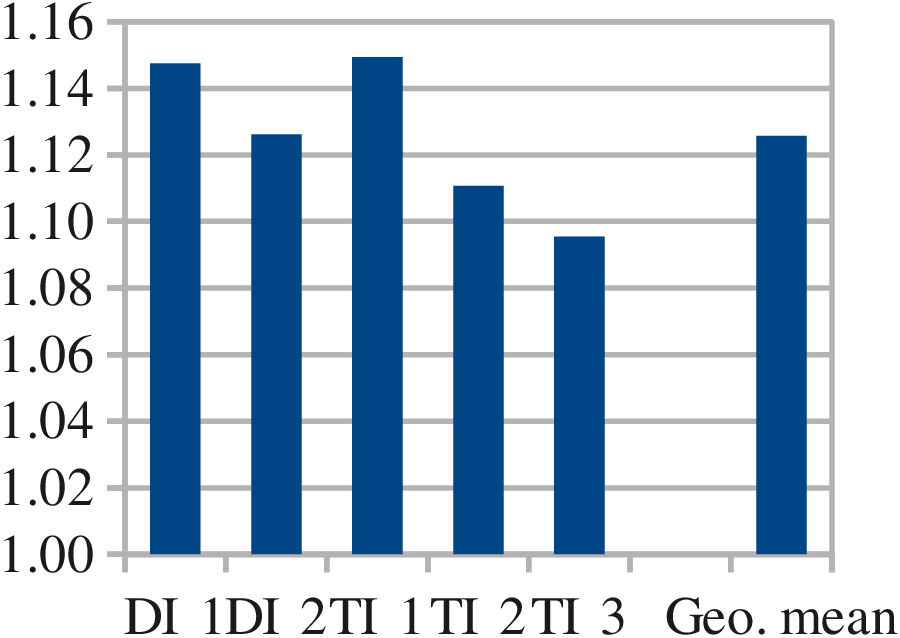}}

    \caption{Energy, performance and area of \io{} vs \ooo{} core designs,
    simulating the Streamcluster benchmark.}
    \label{fig:simres-io-vs-ooo}
\end{figure}

In terms of energy, in general, there is no surprise that pipelines with more
resources consume more energy, even if they may be faster. However, there are
interesting comparisons between \textit{equivalent} \io{} and
\ooo{} cores. Here, the term \textit{equivalent} means that cores
have similar configurations, except the dynamic scheduling capability.
Figure~\ref{fig:auto-tun-simres-io-ooo-area} shows the area overhead of \ooo{}
cores compared to equivalent \io{} designs.

Still analyzing Streamcluster, when the reference kernels execute in equivalent \io{} cores,
their performance is worsened in average by 16~\%, yet being 21~\% more energy
efficient. 
On the other
hand, online auto-tuning improves those numbers to
6~\% and 31~\%, respectively.
In other words, the online approach can considerably
reduce the performance gap between \io{} and \ooo{} pipelines to
only 6~\%, and further improve energy efficiency.

It is also interesting to compare reference kernels executed in
\ooo{} cores to online auto-tuning versions executed in
equivalent \io{} ones. Despite the clear hardware
disadvantage, the run-time approach can still provide average speedups of 1.52 and
1.03 for SISD and SIMD, and improve the energy efficiency by 62~\% and 39~\%,
respectively, as Figure~\ref{fig:auto-tun-simres-io-at-ooo-ref} illustrates.

In the simulations of VIPS, the memory-boundedness is even more
accentuated, because the benchmark is only called once and then Linux does not
have the chance to use disk blocks cached in RAM. The performance of the proposed
approach virtually matched those of the reference kernels. The speedups
oscillate between 0.98 and 1.03, and the geometric mean is 1.00. Considering
that between 29 and 79 new kernels were generated and evaluated during the
benchmark executions, this demonstrates that the proposed technique has negligible
overheads if auto-tuning can not find better versions.

\subsection{Analysis with varying workload}

To better illustrate the behavior of the online auto\hyp{}tuning framework, we
further analyzed its behavior in the CPU\hyp{}bound benchmark, with varying size
of input set and hence execution time. The dimension was varied between 4 and
128 (the native dimension), and the workload through the number of points
between 64 and 4096 (that of \mbox{\texttt{simsmall}}). The other parameters
were set as those in the \mbox{\texttt{simsmall}} input set.

\begin{figure}
    \centering
    \subfigure[Varying dimension in the Cortex\hyp{}A8.\label{fig:auto-tun-a8-dim}]{
        \includegraphics[width=0.48\columnwidth]{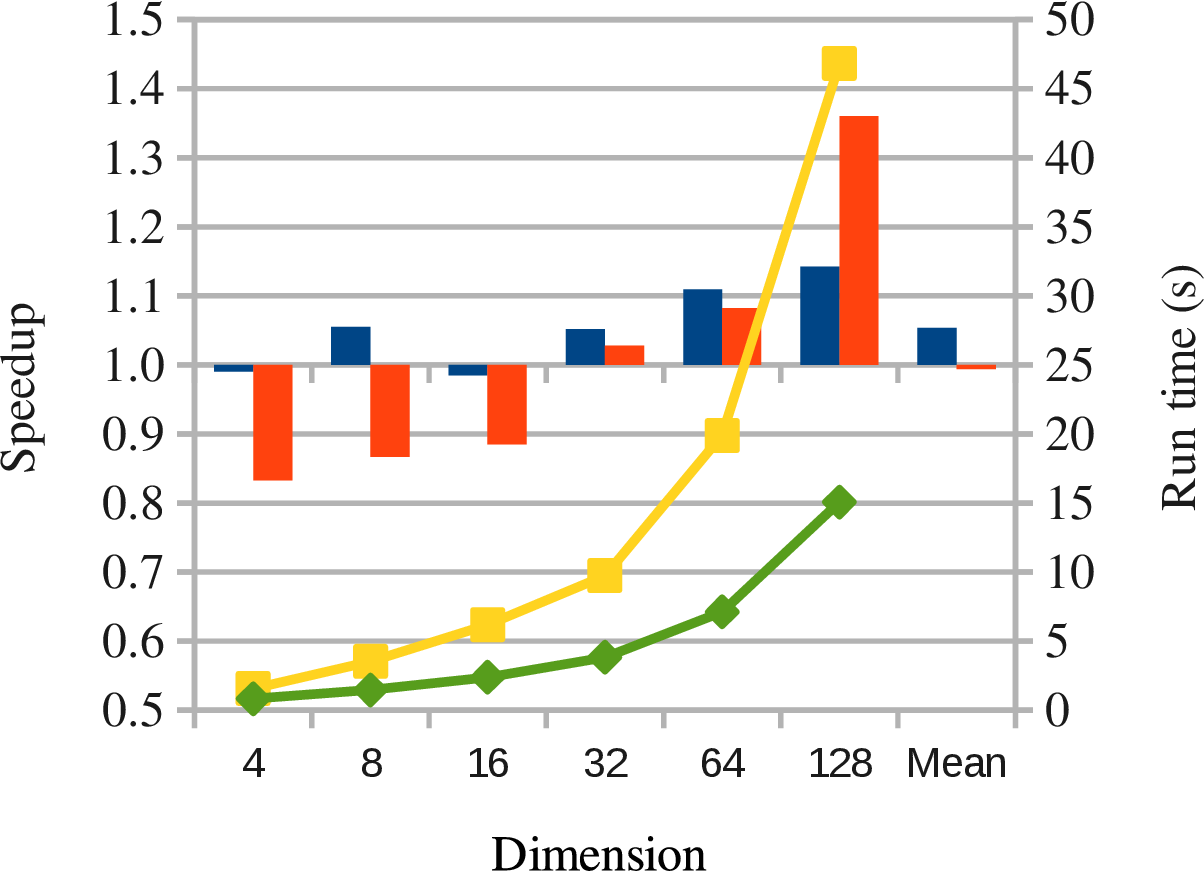}}
    \subfigure[Varying dimension in the Cortex\hyp{}A9.\label{fig:auto-tun-a9-dim}]{
        \includegraphics[width=0.48\columnwidth]{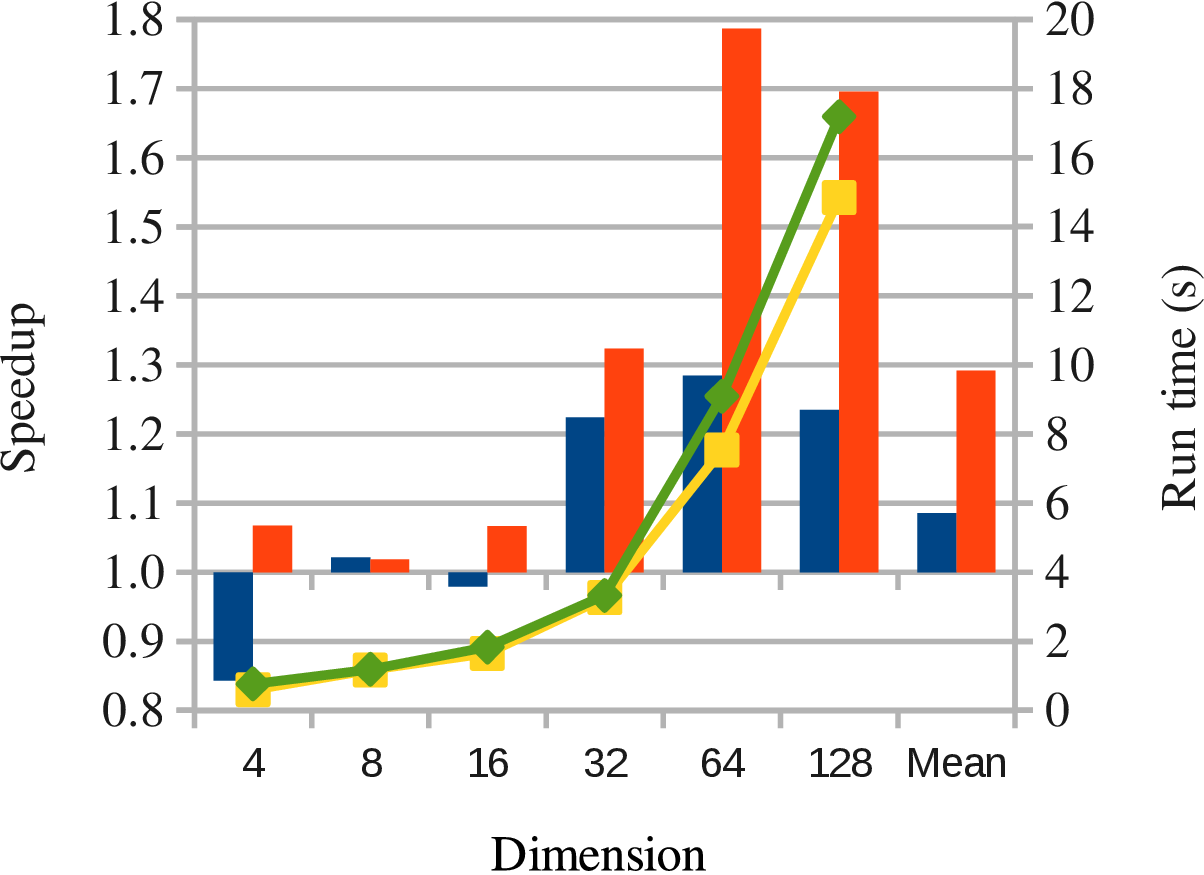}}
    \subfigure[Varying workload in the Cortex\hyp{}A8.\label{fig:auto-tun-a8-points}]{
        \includegraphics[width=0.48\columnwidth]{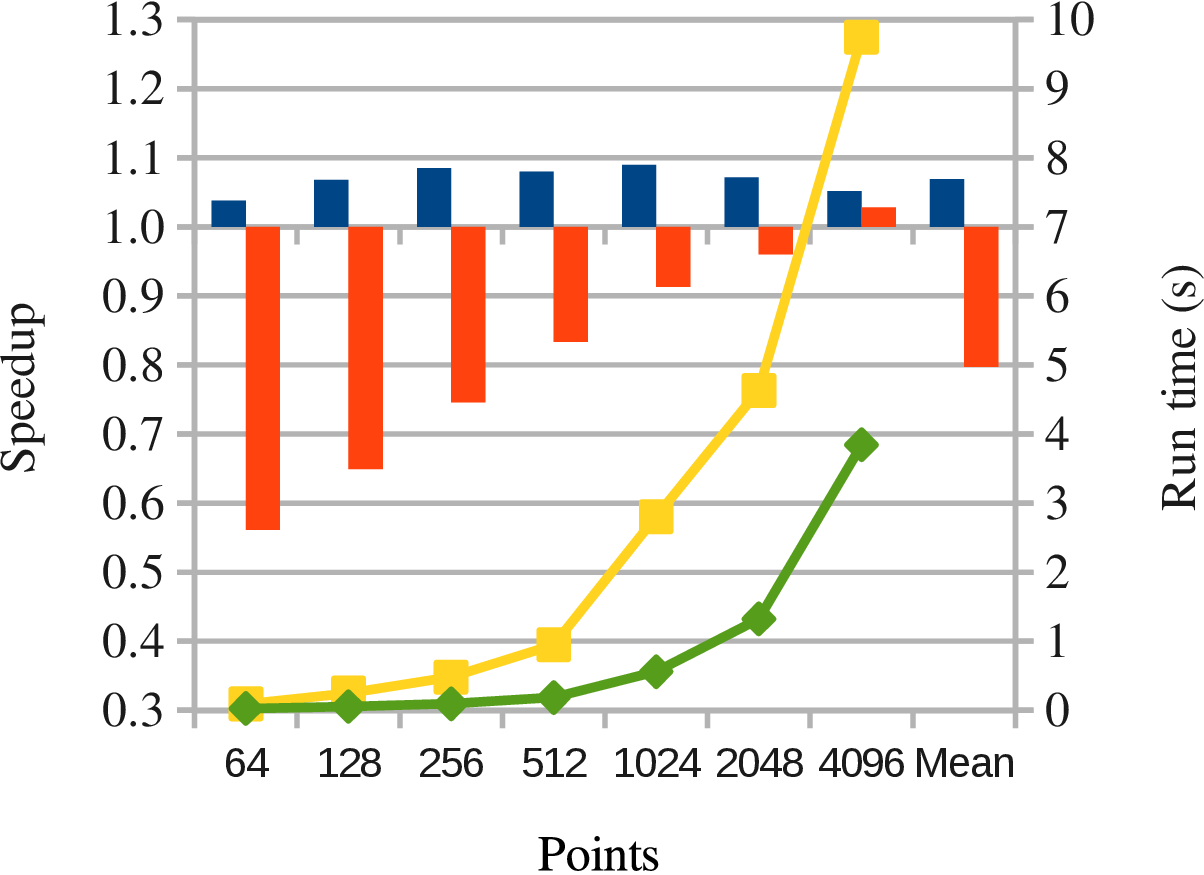}}
    \subfigure[Varying workload in the Cortex\hyp{}A9.\label{fig:auto-tun-a9-points}]{
        \includegraphics[width=0.48\columnwidth]{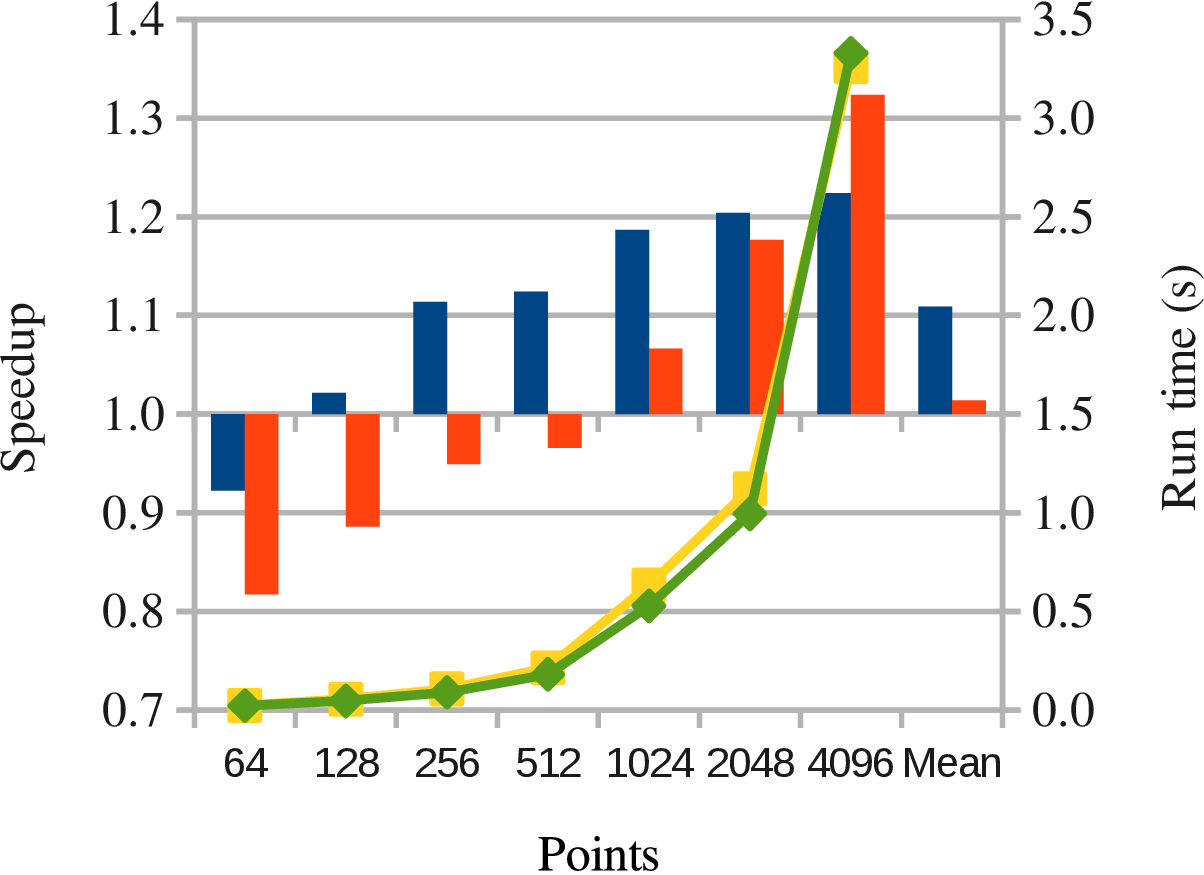}}
    \includegraphics[width=0.48\columnwidth]{auto-tun-a8-points-crop.png}
    \caption{Analysis of the online auto\hyp{}tuning speedups in Streamcluster
        with varying dimension and workload (though the number of points in the
        space), compared to the static references.
        Other parameters are those from the \mbox{\texttt{simsmall}} input set.}
    \label{fig:auto-tun-dim-points}
\end{figure}

Figure~\ref{fig:auto-tun-dim-points} shows all the results in the two real
platforms. Globally, the online auto\hyp{}tuning framework can find the right
balance of space exploration for SISD code, in average obtaining speedups
between 1.05 and 1.11, in applications that run for tens of milliseconds to tens
of seconds. On the other hand, for SIMD auto\hyp{}tuning, average speedups go
from 0.80 to 1.29, with considerable slowdowns in the A8 when the applications
run during less than one second.

Figures~\ref{fig:auto-tun-a8-dim} and \ref{fig:auto-tun-a8-points} show the
speedups obtained in the Cortex\hyp{}A8. We observe that SISD auto\hyp{}tuning
has almost always positive speedups, but SIMD auto\hyp{}tuning shows
considerable slowdowns with small workloads. There are two facts that combined
explain these
slowdowns: the initial active function is the SISD reference code, and in the
Cortex\hyp{}A8, SISD FP instructions execute in the non\hyp{}pipelined VFP
extension, but the SIMD ones execute in the pipelined NEON unit. Therefore,
because the benchmark starts executing SISD code from PARSEC and the
reference run\hyp{}time comes from the SIMD code from PARVEC, with small
workloads, online auto\hyp{}tuning can not pay off.

Figures~\ref{fig:auto-tun-a9-dim} and \ref{fig:auto-tun-a9-points} show the same
analysis in the Cortex\hyp{}A9. In this core, the VFP and NEON units are both
pipelined. In consequence, the considerable slowdowns observed in the A8 with SIMD
auto\hyp{}tuning does not happen. In average, run\hyp{}time auto\hyp{}tuning
provides positive speedups. As in the A8, SISD auto\hyp{}tuning almost
always results in positive speedups. SIMD auto\hyp{}tuning shows slowdowns with
small workloads, but after its crossover around 500~ms, speedups go up to almost 1.8.

\subsection{Analysis of correlation between auto\hyp{}tuning parameters and pipeline
designs}

\begin{figure}
    \centering
    \subfigure[Abscissa: from simpler to more complex cores.]{
	\includegraphics[width=0.48\textwidth]{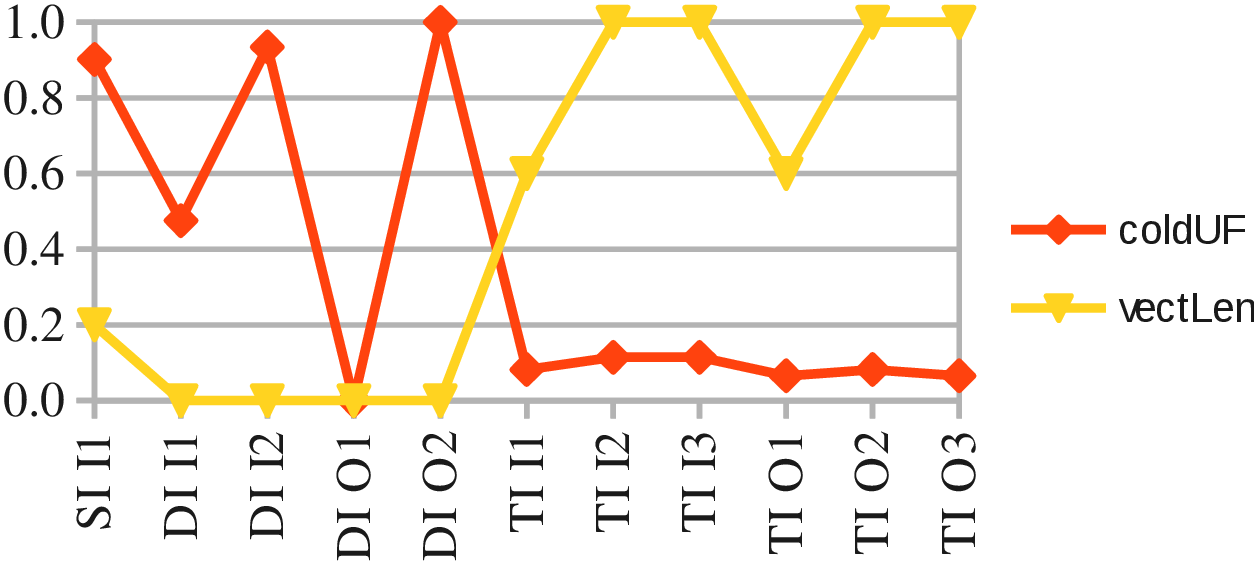}}
    \subfigure[Abscissa: from simpler to more complex cores, \io{} first.]{
	\includegraphics[width=0.48\textwidth]{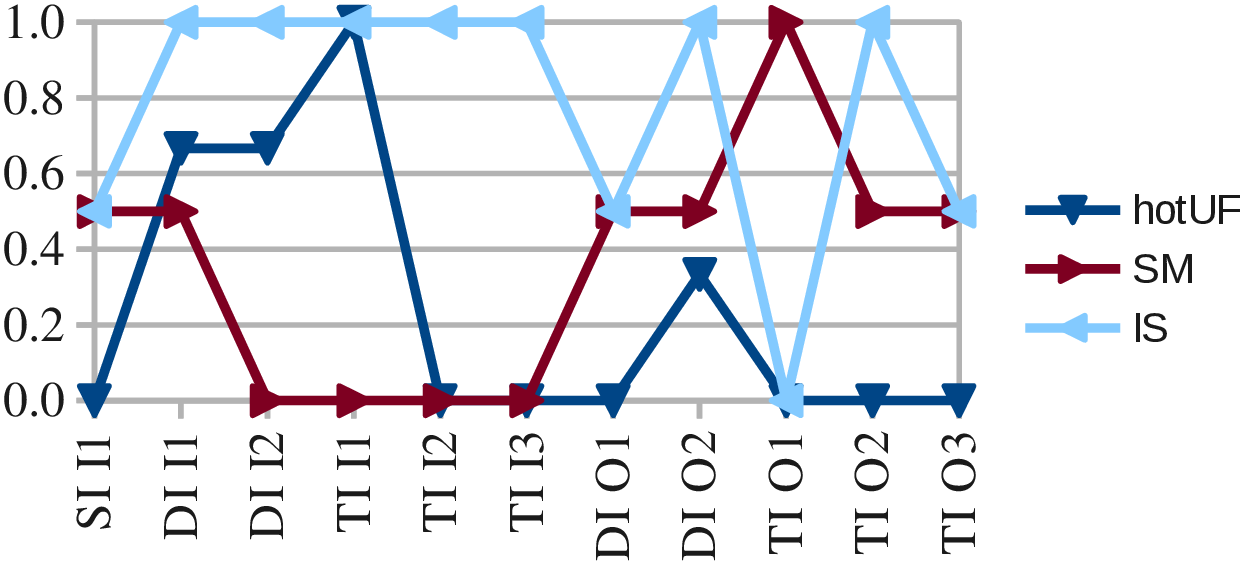}}
    \caption{Normalized values of the averaged best auto\hyp{}tuning parameters for the Streamcluster benchmark, in the 11 simulated cores. \mbox{\texttt{pldStride}} is not shown. Core abbreviations are listed in Table~\ref{tab:auto-tun-sim-core-abbr}.}
    \label{fig:auto-tun-param-matching-norm}
\end{figure}

Table~\ref{tab:auto-tun-param-matching} shows the average auto\hyp{}tuning
parameter values of the best kernel configurations dynamically found in the 11
simulated cores, running the Streamcluster benchmark.
Figure~\ref{fig:auto-tun-param-matching-norm} presents them in the normalized
range from 0 to 1. By analyzing the statistics, it is possible to correlate some
of the most performing parameters to the pipeline designs:

\begin{itemize}
    \item \mbox{\texttt{hotUF}}: This parameter loosely correlates with the dynamic
        scheduling capability of the pipeline. Given that it corresponds to the
        loop unrolling factor without register reuse, balanced \ooo{} cores do
        not benefit from it, because register renaming does the same in hardware
        and allocating more registers can increase the stack and register
        management. Then, 3 of the 4 cores where \mbox{\texttt{hotUF}} was not 1 are \io{}
        designs.
    \item \mbox{\texttt{coldUF}}: This parameter correlates with pipeline depth.
        It corresponds to the loop unrolling factor with register reuse, and
        benefits shallow execution stages. This happens probably because of three
        factors: the dynamic instruction count (DIC) is reduced when a loop is
        unrolled, \mbox{\texttt{coldUF}} is the only parameter that allows aggressive unrolling
        factors, and deeper pipelines need more ILP than reduced DIC. In
        consequence, higher \mbox{\texttt{coldUF}} values are found in single- and
        dual\hyp{}issue designs.
    \item \mbox{\texttt{vectLen}}: It correlates with the pipeline width. This
        parameter defines the length of processing vectors in the loop body, and
        enables higher ILP. That is why triple\hyp{}issue designs have
        $\mathrm{\mbox{\texttt{vectLen}}} \geq 3$, while narrower pipelines have it around 2.
    \item \mbox{\texttt{pldStride}}: It has no clear correlation, possibly because all
        cores have stride prefetchers and the same L1 cache line length.
    \item Stack minimization (SM): This code generation option has a
        loose correlation with the dynamic scheduling capability of the
        pipeline. Even with fewer architectural registers available for
        allocation, \ooo{} designs can still get rid of false
        register dependencies by renaming architectural registers. The reduced
        stack management can then speed up execution.
    \item Instruction scheduling (IS): All types of pipeline benefit
        from instruction scheduling. \OOO{} designs may
        sometimes not need scheduling at all, as we observe in 3 of 5 \ooo{}
        cores, whose average utilization of IS was not 1.
\end{itemize}

\begin{table}
\begin{center}
\begin{threeparttable}[b]
  \caption{Average of the best auto\hyp{}tuning parameters for the Streamcluster
benchmark, in the 11 simulated
cores. Between parenthesis, the parameter ranges are shown.}
  \label{tab:auto-tun-param-matching}
  \begin{tabular}{|l|c|c|c|c|c|c|} \hline
    \monocol{|c|}{\multirow{2}{*}{Core}} & \mbox{\texttt{hotUF}} & \mbox{\texttt{coldUF}} & \mbox{\texttt{vectLen}} & \mbox{\texttt{pldStride}}   & SM     & IS  \\
                                         & (1-4) & (1-64) & (1-4)   & (0, 32, 64) & (0, 1) & (0, 1) \\ \hline
    SI-I1 & 1.0  & 11.7 & 2.3  & 21   & 0.2  & 0.8  \\ \hline
    DI-I1 & 1.3  & 7.3  & 2.0  & 43   & 0.2  & 1.0  \\ \hline
    DI-I2 & 1.3  & 12.0 & 2.0  & 37   & 0.0  & 1.0  \\ \hline
    DI-O1 & 1.0  & 2.5  & 2.0  & 27   & 0.2  & 0.8  \\ \hline
    DI-O2 & 1.2  & 12.7 & 2.0  & 37   & 0.2  & 1.0  \\ \hline
    TI-I1 & 1.5  & 3.3  & 3.0  & 48   & 0.0  & 1.0  \\ \hline
    TI-I2 & 1.0  & 3.7  & 3.7  & 48   & 0.0  & 1.0  \\ \hline
    TI-I3 & 1.0  & 3.7  & 3.7  & 43   & 0.0  & 1.0  \\ \hline
    TI-O1 & 1.0  & 3.2  & 3.0  & 37   & 0.3  & 0.7  \\ \hline
    TI-O2 & 1.0  & 3.3  & 3.7  & 21   & 0.2  & 1.0  \\ \hline
    TI-O3 & 1.0  & 3.2  & 3.7  & 27   & 0.2  & 0.8  \\ \hline
  \end{tabular}
\end{threeparttable}
\end{center}
\end{table}

In this study, we observed correlations between auto\hyp{}tuning parameter
and pipeline features.
The results corroborate the capability of the
auto-tuning system to adapt code to different micro-architectures.
On the other hand, precise correlations could not be identified, because the
best auto-tuning parameters depend on several factors (system load,
initial pipeline state, cache behavior, application phases, to name a few),
whose behaviors can not be easily modeled in complex systems.  Online
auto-tuning is a very interesting solution in this scenario.

\section{Related work}
\label{sec:related}

ADAPT~\cite{Voss:2000:AAD:850941.852890} was one of the first frameworks to
dynamically optimize a running application, without prior user intervention. The
dynamic compilation was performed in an external processor or in a free
processor inside the computing system. In three experiments, ADAPT obtained
speedups between 1.00 and 1.40 in workloads that run during more than one hour
on an uniprocessor.

Active Harmony~\cite{Tiwari:2011:OAC:2058524.2059525} is a run-time compilation
and tuning framework for parallel programs.  By defining a new language,
programmers can express tunable parameters and their range, which are explored
at run time.  In experiments performed in cluster of computers, the code
generation is deployed in idle machines, in parallel to the running application.
Active Harmony provided average speedups between 1.11 and 1.14 in two scientific
applications, run with various problem sizes and in three different platforms.
In their experiments, the minimum time required to obtain speedups was more than
two minutes running in a cluster with 64 cores.

IODC~\cite{Chen:2012:IOD:2150976.2150983} is a framework of iterative
optimization of data center programs within and across runs, which is
transparent to the user. The key idea is to control the intensity of space
exploration (recompilations and runs) accordingly to the savings achieved since
the beginning of the iterative process. In six compute\hyp{}intensive workload,
IODC achieved an average speedup of 1.14, after thousands of runs.

SiblingRivalry~\cite{Ansel:2012:SOA:2380403.2380425} is an auto\hyp{}tuning
technique in which half of cores in homogeneous multicore processors are used to explore
different auto\hyp{}tuning possibilities with online learning, while the other
half run the best algorithm found so far.  It can both adapt code
on\hyp{}the\hyp{}fly to changing load in the system, and to migrations between
micro\hyp{}architectures.  In this latter case, after running 10 minutes, in
average this approach both speeded up by 1.8 and reduced the energy consumption
of eight benchmarks by 30~\%, compared to statically auto-tuning code for Intel
Xeon and running in AMD Opteron, and vice-versa (reference versions were allowed
to use all cores).

UNIDAPT is a framework that enables dynamic optimizations through static
multi\hyp{}versioning~\cite{6495999}, and it is part of the Collective Mind
Framework~\cite{Fursin2014}. A small set of pre\hyp{}optimized versions of a
code together with online learning allow to quickly select the most appropriate
versions at run\hyp{}time, predicting or reacting to changing underlying
(heterogeneous) architectures or varying program phases.

Previous work proposed online auto\hyp{}tuning in DSCPs, usually regenerating
code in idle cores or machines. ADAPT, IODC and SiblingRivalry required several
minutes, hours or runs to obtain performance gains over static compilation.
Active Harmony obtained speedups after only two minutes, but using idle
cores in a cluster of computers. Instead, our work is the first to focus on
embedded processors, implementing online auto\hyp{}tuning in short\hyp{}running
kernels, and obtaining positive speedups after hundreds of milliseconds, without
extra cores to regenerate code and including all overheads.
UNIDAPT has also a very low run\hyp{}time overhead, because good code versions
are statically found through iterative compilation, however with radical
changes of micro\hyp{}architecture or input data size, static multi\hyp{}versioning
may not be able to obtain the desired performance with a small number of
versions or may lead to code size explosion.
The average speedups achieved by our
framework (1.16 in 2 real cores and 1.17 in 11 simulated cores) in 12 benchmark
configurations (2 benchmarks, 3 input sets, SISD and SIMD) are comparable to
those obtained by previous work, except SiblingRivalry that achieved higher
speedups, but in a different experimental setup.
Furthermore, our approach is scalable to heterogeneous multi\slash{}manycores, because
the same computing kernel running in distinct threads can be locally
auto\hyp{}tuned to each core configuration, thanks to the low overheads per core.

In addition, in a detailed simulation experiment of 60 running configurations
(10 similar \io{} and \ooo{} CPUs running 1 benchmark with 3 input sets, SISD and
SIMD), we demonstrated that our online approach can virtually replace hardware
out\hyp{}of\hyp{}ordering, running a highly CPU\hyp{}bound benchmark in \io{}
designs, still widely deployed in the embedded market. These are new results,
and to the best of our knowledge, no previous work tried to elucidate and
quantify this possibility.

\section{Conclusion} \label{sec:auto-tun-conclusion}

In this paper, we presented an approach to implement run-time
auto-tuning kernels in short-running applications.
This work advances the state of the art of online auto-tuning. To the best
of our knowledge, this work is the first to propose an approach of online
auto-tuning that can obtain speedups in short-running kernel-based applications. Our
approach can both adapt a kernel implementation to a micro-architecture
unknown prior compilation and dynamically explore auto-tuning possibilities
that are input-dependent.

We demonstrated
through two case studies in real and simulated platforms that the proposed approach can
speedup a CPU-bound kernel-based application up to 1.79
and 2.53, respectively, and has negligible run-time overheads when
auto-tuning does not provide better kernel versions. In the second
application, even if the bottleneck is in the main memory, we observed speedups
up to 1.30 in real cores, because of the reduced number of instructions executed
in the auto-tuned versions.

Energy consumption is the most constraining factor in current
high-performance embedded systems. By simulating the CPU-bound
application in 11 different CPUs, we showed that run-time auto-tuning can reduce
the performance gap between \io{} and \ooo{} designs from 16~\%
(static compilation) to only 6~\%. In addition, we demonstrated that online
micro-architectural adaption of code to \io{} pipelines can in
average outperform the hand vectorized references run in similar
\ooo{} cores. Despite the clear hardware disadvantage, online
auto-tuning in \io{} CPUs obtained an average speedup of 1.03 and an
energy efficiency improvement of 39~\% over the SIMD reference in \ooo{} CPUs.

\section*{Acknowledgments}
This work has been partially supported by the LabEx PERSYVAL-Lab
(ANR-11-LABX-0025-01) funded by the French program Investissement d'avenir.

\bibliographystyle{plain}
\bibliography{Fernando_}

\end{document}